\documentclass[aps,pre,reprint,10pt,superscriptaddress,showpacs]{revtex4-2}

\usepackage{amsfonts} 
\usepackage{amsmath}
\usepackage{amssymb}
\usepackage{subfigure}
\usepackage{graphicx}
\usepackage{dcolumn}
\usepackage{bm}
\usepackage{textcomp}
\usepackage{amsmath}
\usepackage{verbatim}
\usepackage{comment}
\usepackage{lipsum}
\usepackage{subfigure}
\usepackage{color}
\usepackage{soul}
\usepackage{cancel}
\usepackage{ulem}

\begin{document}
\title[]{Nonlinear modulation of dispersive fast magnetosonic waves in an inhomogeneous rotating solar low-$\beta$ magnetoplasma}
\author{Jyoti Turi}
\email{ jyotituri.maths@gmail.com}
\affiliation{Department of Mathematics, Siksha Bhavana, Visva-Bharati University, Santiniketan-731 235, West Bengal, India}
\author{A. P. Misra}
\homepage{Author to whom any correspondence should be addressed}
\email{apmisra@visva-bharati.ac.in}
\affiliation{Department of Mathematics, Siksha Bhavana, Visva-Bharati University, Santiniketan-731 235, West Bengal, India} 
\begin{abstract}
	We study the modulation of fast magnetosonic waves (MSWs) in rotating inhomogeneous low-$\beta$ magnetoplasmas with the effects of gravitation and the Coriolis force. By employing the standard multiple-scale reductive perturbation technique (RPT), we derive a nonlinear Schr\"{o}dinger (NLS) equation that governs the evolution of slowly varying MSW envelopes. The fast MSW  becomes dispersive by the effects of the Coriolis force in the fluid motion, and the magnetic field and density inhomogeneity effects favor the Jeans instability in self-gravitating plasmas in a larger domain of the wave number ($k$,  below the Jeans critical wave number, $k_J$) than homogeneous plasmas. The relative influence of the Jeans frequency ($\omega_J$, associated with the gravitational force) and the angular frequency ($\Omega_0$, relating to the Coriolis force) on the Jeans carrier MSW mode and the modulational instability (MI) of the MSW envelope is studied. We show that the MSW envelope (corresponding to the unstable carrier Jeans mode with $\omega_J>2\Omega_0$ and $k<k_J$) is always unstable against the plane wave perturbation with no cut-offs for growth rates.  In contrast, the stable Jeans mode with $\omega_J>2\Omega_0$ but $k>k_J$ manifests either modulational stability or MI having a finite growth rate before being cut off. We find an enhancement of the MI growth rate by the influence of magnetic field or density inhomogeneity. The case with constant gravity force (other than the self-gravity) perpendicular to the magnetic field is also briefly discussed to show that the fast magnetosonic carrier mode is always unstable, giving MI of slowly varying envelopes with no cut-offs for the growth rates.  Possible applications of MI in solar plasmas, such as those in the X-ray corona, are also briefly discussed. 
\end{abstract}

\maketitle

\section{Introduction} \label{sec-intro}
One-fluid magnetohydrodynamics deals with a compressible conducting fluid immersed in a magnetic field and it is often regarded
as a reasonable description of the large-scale dynamics of a plasma. Magnetohydrodynamic (MHD) waves are generally described applying this theory and represent one of the macroscopic processes responsible for the transformation of energy and information in plasmas. The theory of MHD waves in an infinite conducting medium was first developed by Alfv{\'e}n \cite{alfven1942existence} with its application to sunspots, coronal heating, particle acceleration, and generation of cosmic radiation. Since then, the study of nonlinear MHD waves in plasma has been one of the most popular research topics among researchers, given its remarkable application and progress in laboratory experiments and techniques as well as various space and astrophysical plasma environments like pulsar magnetosphere, magnetars, solar corona, etc. 
\par
 Magnetosonic waves (MSWs) are one of the fundamental MHD wave modes in plasmas, which propagate nearly perpendicular to the background magnetic field and are often observed in laboratory
plasmas \cite{hannan2013fast}, earth magnetosphere \cite{stasiewicz2000small,stasiewicz2003slow,stasiewicz2004theory,shukla2004nonlinear,shukla2011alfvenic}, and solar wind plasmas \cite{klein2012using}, etc.
These waves are of great interest because of their important roles in plasma heating \cite{schmidt2011slow} and charged particle acceleration \cite{rau1998strongly}. Several authors have investigated the characteristics of MSW waves in different space and astrophysical plasma environments. To mention a few, Marklund \textit{et al.} \cite{marklund2007magnetosonic} studied the magnetosonic solitons in a quantum magnetoplasma, including the quantum Bohm
potential and electron spin-1/2 effects using the Sagdeev
potential approach. Haas and Mahmood \cite{haas2018magnetosonic}  analyzed the propagation of linear
and weakly nonlinear magnetosonic waves in a plasma with
arbitrary degeneracy of electrons and with the inclusion of Bohm
diffraction effects. Nonlinear properties of fast MSWs in dense dissipative plasmas with degenerate electrons were studied theoretically by Masood \textit{et al.}  \cite{masood2014nonlinear}  by deriving the Zabolotskaya-Khokhlov (ZK) equation for small but finite amplitude excitations. Hussain and Mahmood  \cite{hussain2011korteweg} investigated the nonlinear propagation of MSWs and showed that these waves may evolve into shock-like structures that may be responsible for heating
the solar chromosphere and the solar corona.
  \par 
  The modulational instability (MI) of nonlinear waves in dispersive or diffractive media has been known to be one of the most important mechanisms of energy localization via the formation of different localized coherent structures like envelope solitons \cite{salahuddin2002ion}, envelope shocks \cite{sultana2012electron}, freak wave (or rouge waves) \cite{mckerr2014freak,chowdhury2017rogue}, giant waves \cite{ruderman2010freak}, etc., as well as the transfer of energy between waves and particles, leading to particle heating, e.g., heating of coronal loops \cite{habbal1979}.
In this context,
Watanabe \cite{watanabe1977self} first experimentally observed the MI of nonlinear wave envelopes in dispersive media in $1977$. Subsequently, the investigations of MI and associated nonlinear structures have gained significant attention among researchers, and a large number of theoretical and experimental investigations on MI of electrostatic and electromagnetic waves have been done to explain the effects of different physical parameters in various space and astrophysical plasmas \cite{sultana2011electrostatic, shalini2015modulation, misra2007nonlinear,bains2010modulational}.
 For example, Sahyouni \textit{et al.}  \cite{sahyouni1988dark} studied the amplitude modulation of fast magnetosonic surface waves in solar flux tubes and showed that the fast wave-envelope admits dark envelope solitons solution and discussed the possibility of the existence of solitary waves in the solar atmosphere.
 Sakai \cite{sakai1983modulational} examined the  MI of fast MSWs theoretically and discussed its applications to solar plasmas. Misra and Shukla \cite{misra2008modulational} studied the MI of magnetosonic waves using a two-fluid quantum magnetohydrodynamic model that includes the effects of the electron-1/2 spin and the plasma resistivity.
Panwar \textit{et al.} have investigated the MI and associated rogue-wave  structures of slow magnetosonic perturbations in a Hall-MHD plasma \cite{panwar2014modulational}. 
Wang \textit{et al.} have studied the MI of MSWs in a Fermi-Dirac-Pauli plasma by the combined effects of the electron relativistic degeneracy, the quantum tunneling, electron spin via Pauli paramagnetism, and plasma resistivity \cite{wang2013modulational}.
  \par
 To the best of our knowledge, the combined effects of the Coriolis force and the gravity force on the fast magnetosonic modes and the MI of slowly varying envelopes in inhomogeneous magnetoplasmas have not been studied before. Our aim is to study the propagation characteristics of fast magnetosonic modes and their nonlinear evolution as slowly varying envelopes through the modulational instability. To this end, we consider a self-consistent MHD model for electron and ion fluids that includes the combined effects of the gravity and Coriolis forces as well as the magnetic field and density inhomogeneities in absence of any viscosity or magnetic diffusivity. We show that while the fast magnetosonic modes  exhibit instability in plasmas with constant gravity perpendicular to the magnetic field, the fast magnetosonic Jeans mode in self-gravitating plasmas can be stable or unstable depending on the competitive roles of the gravity and Coriolis  forces and whether the Jeans wave number is below or above a critical value. The possibility of the emergence of MI of slowly varying magnetosonic envelopes  in different cases of the stable or unstable fast carrier modes in presence of the self-gravitation and the constant gravity force with the effects of magnetic field or density inhomogeneities are studied and its applications to solar coronal plasmas are discussed. 
  \par The paper is organised in the following fashion: In Sec. \ref{sec-bas-eq}, we present the basic MHD models governing the dynamics of fast magnetosonic waves and the inhomogeneous equilibrium state of plasmas. Section \ref{sec-nls-sg} demonstrate the linear fast magnetosonic mode, the compatibility condition, and the derivation of the nonlinear Schr{\"o}dinger (NLS) equation for the evolution of slowly varying magnetosonic envelopes.  We study the MI in Sec. \ref{sec-MI} for three different cases of stable and unstable Jeans carrier modes as well as for the unstable carrier mode under the influence of gravity.  The applications of our results in solar plasmas are discussed in Sec. \ref{sec-appl}. Finally, Sec. \ref{sec-summary} is left to summarize and conclude our results.   
\section{Basic equations} \label{sec-bas-eq}
\begin{figure*}
	\centering
	\includegraphics[width=0.6\textwidth]{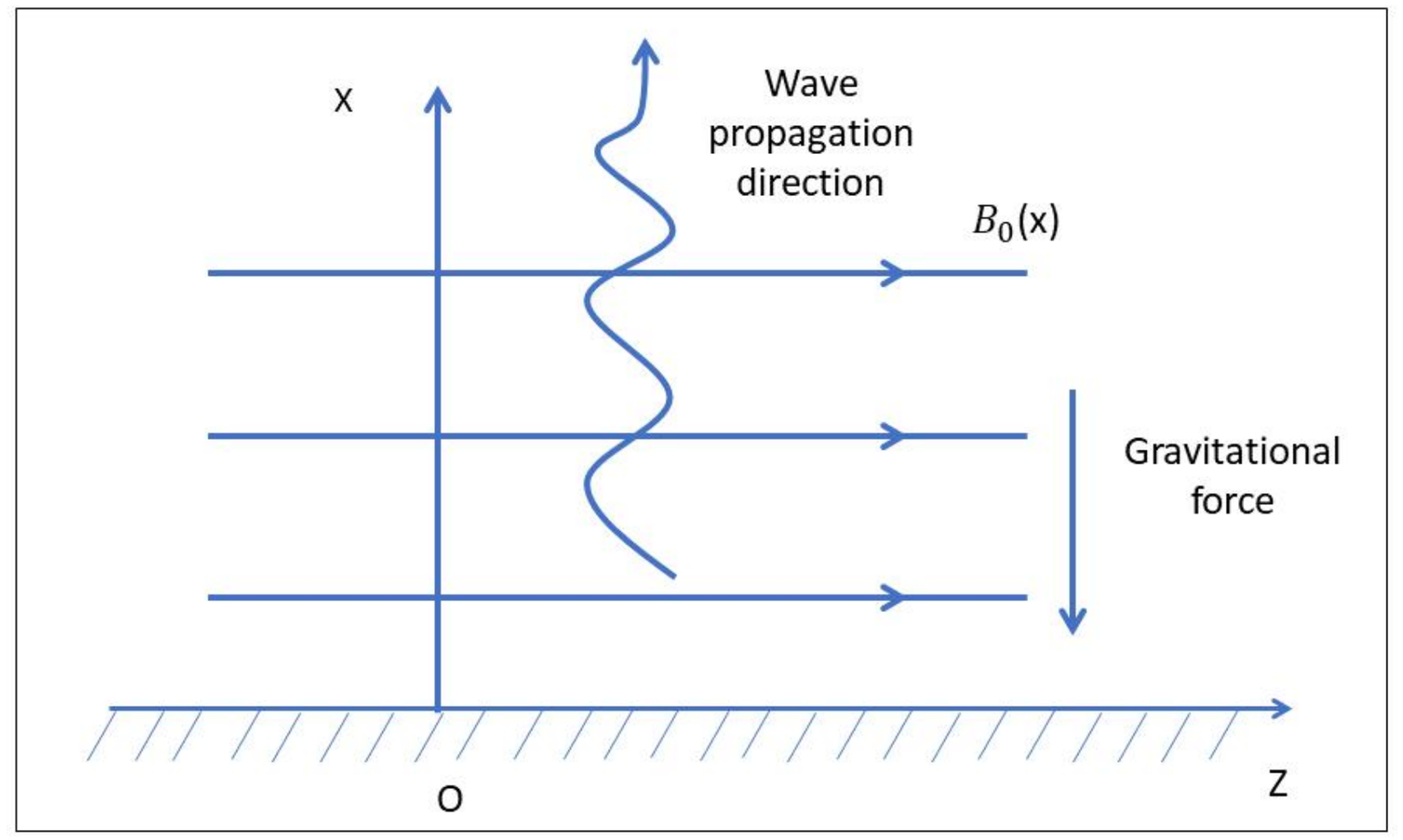}
	\caption{A schematic diagram showing the directions of wave propagation $\mathbf{k}$, the external magnetic field $\mathbf{B_0}$, and the gravitational force $\mathbf{g}$. The background plasma state is determined by Eqs. \eqref{eq-P0}, \eqref{eq-psi0}. The magnetic field and density inhomogeneities are along the $x$-axis.}
	\label{fig0-diag}
\end{figure*}

We consider the nonlinear propagation of fast magnetosonic waves in a rotating magnetized plasma with the effects of the Coriolis force and the gravitational force. The latter may be considered in two cases (i) when the gravitational acceleration $g$ is not a constant (self-gravity) and (ii) when $g$ is a constant.  The plasma is supposed to be rotating with uniform angular velocity $\mathbf{\Omega}=(0,\Omega_0 \cos \lambda,\Omega_0 \sin\lambda)$ and immersed in an external static magnetic field along the $z$-axis, i.e.$\mathbf{B_0}=B_0 \hat{z}$. We assume the gravitational force to be acting vertically downwards and parpendicular to the magnetic field and the wave propagation along the $x$-axis, for simplicity. Furthermore, the background magnetic field and the density are assumed to vary along the $x$-axis (inhomogeneities). A schematic diagram for the model is  shown in Fig. \ref{fig0-diag}. We first consider the case of self-gravitating plasmas. 
The basic equations describing the dynamics of electron-ion fluids in self-gravitating magnetoplasmas in the center-of-mass frame are \cite{turi2022magnetohydrodynamic}  
\begin{equation}
	\label{eq-cont}
	\frac{\partial \rho}{\partial t} +\mathbf{\nabla}\cdot\left(\rho \mathbf{v}\right)=0, 
\end{equation} 
\begin{multline}
	\label{eq-moment}
	\frac{\partial \mathbf{v}}{\partial t} +\left(\mathbf{v}\cdot\mathbf{\nabla}\right) \mathbf{v}=-\frac{1}{\rho}\mathbf{\nabla}\left(P+\frac{\mathbf{B}^2}{2\mu_0}\right)
	\\+\frac{1}{\rho \mu_0} \left(\mathbf{B}\cdot\mathbf{\nabla}\right)\mathbf{B}-2\mathbf{\Omega}\times\mathbf{v} +\mathbf{g}, 
\end{multline}

\begin{equation}
	\label{eq-B}
	\frac{\partial \mathbf{B}}{\partial t} +\left(\mathbf{v}\cdot\mathbf{\nabla}\right) \mathbf{B}=\left(\mathbf{\textbf{B}}\cdot\mathbf{\nabla}\right) \mathbf{v}-\left(\mathbf{\nabla}\cdot\mathbf{v}\right) \mathbf{B},
\end{equation}
\begin{equation}
	\label{eq-poiss}
	\mathbf{\nabla}\cdot\mathbf{g}=-4\pi G \rho,
\end{equation}
where $\mathbf{g}=-\mathbf{\nabla}\psi$ is the gravitational force per unit mass of the fluid acting vertically downwards; $\rho$, $\mathbf{v}$, and $P$ are, respectively, the fluid density, fluid velocity, and the thermal pressure. Also, $\mathbf{B}$, $\mathbf{\Omega}$, $\psi$, and $G$, respectively, denote the magnetic field, uniform angular velocity of the rotating fluid, gravitational potential, and the universal gravitational constant. The pressure $P$ satisfies the equation of state: $\mathbf{\nabla}P=c_s^2 \mathbf{\nabla}\rho$, where $c_s=\sqrt{\gamma k_B T_e/m_i}$ is the adiabatic sound speed with $k_B$ denoting the Boltzmann constant, $T_e$ the electron temperature, $\gamma$ the adiabatic index, and $m_i$ the ion mass.
\par 
Next, we consider the one-dimensional propagation of fast MSWs along the $x$-axis and normalize Eqs. \eqref{eq-cont}-\eqref{eq-poiss} according to:
$\rho\rightarrow\rho/\rho_0$, $B\rightarrow B/B_0$,  $\omega\rightarrow\omega/\omega_{ci}$,
 $\left(v_x,v_y,v_z\right)\rightarrow \left(v_x,v_y,v_z\right)/V_A$, ${c}_s\rightarrow c_s/V_A$,    ${\Omega}_0\rightarrow \Omega_0/\omega_{ci}$, ${\omega}_J\rightarrow \omega_J/\omega_{ci}$,  $\psi\rightarrow \psi/V_A^2$. Here, $V_A={B_0}/{\sqrt{\mu_0 \rho_0}}$ is the Alfv{\'e}n speed, $\omega_{J}^2=4\pi G \rho_0$ is the squared Jeans frequency, and $\omega_{ci}=\sqrt{eB_0/m_i}$ is the ion-cyclotron frequency at $x=0$. Furthermore, the space and time coordinates are normalized as   $x\rightarrow x \omega_{ci}/V_A$ and $t \rightarrow t \omega_{ci}$.  
Thus, from Eqs. \eqref{eq-cont}-\eqref{eq-poiss}, after separating the velocity components along the axes and noting that the frozen-in-field condition $\rho/\rho_0=B/B_0$ holds for Eqs. \eqref{eq-cont} and \eqref{eq-B}, we obtain the following  reduced set of normalized equations 
\begin{eqnarray}
	\frac{\partial B}{\partial t} +\frac{\partial }{\partial x}\left(v_x B\right)=0,  \label{eq-B1}
\end{eqnarray}
\begin{eqnarray}
	\label{eq-moment-vx}
	\frac{\partial v_x}{\partial t} +v_x \frac{\partial v_x}{\partial x} =-{c}_s^2\frac{1}{B} \frac{\partial B}{\partial x} -\frac{1}{2}\frac{1}{B} \frac{\partial B^2}{\partial x} \nonumber \\
	-2\left({\Omega}_{0} v_z \cos\lambda -{\Omega}_{0}v_y \sin\lambda \right)+ \frac{\partial \psi}{\partial x},
\end{eqnarray}
\begin{eqnarray}
	\frac{\partial v_y}{\partial t} +v_x \frac{\partial v_y}{\partial x}=-2 {\Omega}_{0}v_x \sin\lambda, \label{eq-moment-vy}
\end{eqnarray}
\begin{eqnarray}
\frac{\partial v_z}{\partial t}+v_x\frac{\partial v_z}{\partial x}=2 {\Omega}_{0} v_x\cos\lambda,	  \label{eq-moment-vz}
  \end{eqnarray}
\begin{eqnarray}
	\label{eq-poiss1}
	 \frac{\partial^2 \psi}{\partial x^2}+{\omega}_{J}^2B=0.
\end{eqnarray}
It is to be mentioned that in a general manner, Eqs. \eqref{eq-cont}-\eqref{eq-poiss} describe three independent propagating modes, namely, the Alfv{\'e}n wave and the fast and slow magnetosonic waves. These waves can be distinguished by their speeds and polarizations. In the present work, we have chosen the propagation direction along the x-axis, i.e., perpendicular to the external magnetic field $B_0 \hat{z}$. Thus, we have only the fast magnetosonic wave, since the slow magnetosonic wave and the Alfv{\'e}n wave become non-propagating and degenerate. However, if we consider the polarization in a plane and propagation direction along an axis or as arbitrary having components along the axes, there would appear both the Alfv{\'e}n and magnetosonic waves as fundamental wave modes, which may be coupled, e.g., due to magnetic field line curvature \cite{petrashchuk2023}, the Hall effects \cite{ruderman2020} and/or the Coriolis force for rotating fluids \cite{turi2022magnetohydrodynamic}. The nonlinear interactions of these high-frequency Alfv{\'e}n waves and low-frequency magnetosonic waves can give rise to several interesting phenomena, including wave structures \cite{petrashchuk2023} and MHD wave turbulence due to energy exchange \cite{andres2017}. 
\subsection{Equilibrium state} \label{sec-sub-eqbm}
At equilibrium, the background plasma state with density and magnetic field inhomogeneities is defined by the following pressure-balance and the Poisson equations
\begin{equation}
\frac{d}{dx}\left(P_0(x)+\frac{B_0^2(x)}{2\mu_0}\right)=-\rho_0(x)\frac{d \psi_0(x)}{dx}, \label{eq-P0}
\end{equation}
\begin{equation}
\frac{d^2 \psi_0(x)}{dx^2}=4\pi G\rho_0(x), \label{eq-psi0}
\end{equation}
where the suffix `$0$' denotes the equilibrium value of the corresponding physical quantity. We define $B_0(0)=B_0$, $\rho_0(0)=\rho_0$ and $\psi_0(0)=\psi_0$. By means of the normalization defined before,   Eqs. \eqref{eq-P0} and \eqref{eq-psi0} reduce to
\begin{equation}
\frac{d}{dx}\left({c}_s^2\rho_0(x)+\frac{1}{2}B_0^2(x)\right)=-\rho_0(x)\frac{d \psi_0(x)}{dx}, \label{eq-P00}
\end{equation}
\begin{equation}
\frac{d^2 \psi_0(x)}{dx^2}={\omega}_J^2\rho_0(x). \label{eq-psi00}
\end{equation}
Using the relation $\nabla P=c_s^2\nabla\rho$, and noting that $B_0(x)/B_0=\rho_0(x)/\rho_0$ (Frozen-in-field condition), we obtain from Eq. \eqref{eq-P00}, the following relation (in normalized form)
\begin{equation}
c_s^2\ln \rho_0(x)+2\rho_0(x)=\tilde{\psi}_0-\psi_0(x), \label{eq-P01}
\end{equation}
where $\tilde{\psi}_0=2+\psi_0$. Typically, if $\rho_0(x)$ is not too small, i.e., if it lies in $0.5\lesssim\rho_0(x)\lesssim1$, the function $\ln \rho_0(x)/\rho_0(x)$ approaches from small negative values to zero as $\rho_0(x)\rightarrow1$. In addition, if the magnetic force dominates over the pressure gradient force (in low-$\beta$ plasmas such as those in solar corona) or if the length scale of magnetic field inhomogeneity is much smaller than the density inhomogeneity, i.e., $L_{B_0}\ll L_{\rho_0}$, where   $1/L_{\rho_0}\equiv [1/\rho_0(x)](d\rho_0(x)/dx)$ and $1/L_{B_0}\equiv [1/B_0(x)](dB_0(x)/dx)$, then the term proportional to $c_s^2$ in Eq. \eqref{eq-P01} can be neglected compared to the the term involving the factor $2$ (the second term on the left-hand side). Thus, from Eqs.  \eqref{eq-psi00} and \eqref{eq-P01}, we obtain for $x>0$ the following approximate solutions for $\psi_0(x)$ and $\rho_0(x)$.
\begin{equation}
\label{eq-inhom1}
\begin{split}
&\psi_0(x)\approx \tilde{\psi}_0-2\exp\left(-\frac{1}{2}\omega_J^2 x\right),\\
& \rho_0(x)\equiv B_0(x)\approx\frac{1}{2}\left[\tilde{\psi}_0-\psi_0(x)\right].
\end{split}
\end{equation} 
\par 
On the other hand, if the contribution from the pressure gradient force is much higher than the magnetic force (in high-$\beta$ plasmas such as those in the solar photosphere and solar wind acceleration region) or if the length scale of magnetic field inhomogeneity is much larger than the density inhomogeneity, i.e., $L_{B_0}\gg L_{\rho_0}$, approximate solutions for $\rho_0(x)$ and $\psi_0(x)$ can then be obtained from Eqs.  \eqref{eq-psi00} and \eqref{eq-P01} as $(x>0)$
\begin{equation}
\label{eq-inhom2}
\begin{split}
&\psi_0(x)\approx c_s^2\left[1-\bar{\psi}_0\exp\left(-\frac{\omega_j^2}{c_s^2} x\right)\right],\\
&\rho_0(x)\equiv B_0(x)\approx \exp\left(\frac{\psi_0-\psi_0(x)}{c_s^2}\right),
\end{split}
\end{equation}   
where $\bar{\psi}_0=1-\psi_0/c_s^2$. 
 Furthermore, for plasmas where the plasma pressure is comparable to the magnetic pressure $(\beta\sim1)$, such as those in the lower chromospheric region of the solar atmosphere, Eqs.  \eqref{eq-psi00} and \eqref{eq-P01} are to be solved numerically. However, we are not considering these two cases in the present investigation.
%
\section{Derivation of NLS equation} \label{sec-nls-sg}
We study the modulation of weakly nonlinear slowly varying magnetosonic wave envelopes that are generated due to nonlinear self-interactions of carrier fast magnetosonic modes and higher harmonic modes in self-gravitating rotating magnetoplasmas. To this end, we employ the standard multiple-scale reductive perturbation technique (RPT) \cite{asano1969perturbation} to Eqs. \eqref{eq-B1}-\eqref{eq-poiss1} and derive the nonlinear Schr\"{o}dinger equation for the evolution of slowly varying MSW envelopes. In the RPT, we define a new frame of reference in which the space and time variables are stretched as
\begin{equation} 
	\label{eq-stretch}
	\begin{split}
	&\xi =\epsilon\left(x-v_{g} t\right),\\
	&\tau =\epsilon^{2}t, 	
	\end{split} 	
	\end{equation}
where $v_{g}$ is the group velocity of the wave envelope along the $x$-axis and $\epsilon$ is a small ($0<\epsilon\ll1$) expansion parameter, which scales the weakness of amplitudes of perturbations. Due to the stretched coordinates \eqref{eq-stretch}, the space and time derivatives will be replaced according to the following transformations:
\begin{eqnarray}
	\label{eq-diffop}
	\frac{\partial}{\partial t}\rightarrow \frac{\partial}{\partial t}-\epsilon v_g\frac{\partial}{\partial \xi} +\epsilon^{2}\frac{\partial}{\partial \tau},\nonumber \\
	\frac{\partial}{\partial x} \rightarrow \frac{\partial}{\partial x} +\epsilon \frac{\partial}{\partial \xi},\\
	\frac{\partial^2}{\partial x^2} \rightarrow \frac{\partial^2}{\partial x^2} +2 \epsilon \frac{\partial^2}{\partial x \partial \xi} +\epsilon^2 \frac{\partial}{\partial \xi^2}. \nonumber
\end{eqnarray}
\par
 The dynamical variables are divided into unperturbed (equilibrium) and perturbed parts. In the latter, the slow and fast scales (for space and time), respectively, enter  the $l$-th harmonic amplitudes and the phase $(kx-\omega t)$. Thus, the variables can be expanded as
\begin{equation}
		\begin{split}
	&B=B_0(x)+ \sum_{n=1}^{\infty} \epsilon^n \sum_{l=-\infty}^{\infty} B_{l}^{(n)}(\xi, \tau) \exp\left[il\left(kx-\omega t\right)\right],\\
	&v_{x}= 0+\sum_{n=1}^{\infty} \epsilon^n \sum_{l=-\infty}^{\infty} v_{xl}^{(n)}(\xi, \tau) \exp\left[il\left(kx-\omega t\right)\right],  \\
	&v_{y}=0+\sum_{n=1}^{\infty} \epsilon^n \sum_{l=-\infty}^{\infty} v_{yl}^{(n)}(\xi, \tau) \exp\left[il\left(kx-\omega t\right)\right],  \\ 
	&v_{z}=0+ \sum_{n=1}^{\infty} \epsilon^n \sum_{l=-\infty}^{\infty} v_{zl}^{(n)}(\xi, \tau) \exp\left[il\left(kx-\omega t\right)\right], \\
	&\psi=  \psi_0(x)+\sum_{n=1}^{\infty} \epsilon^n \sum_{l=-\infty}^{\infty} \psi_{l}^{(n)}(\xi, \tau) \exp\left[il\left(kx-\omega t\right)\right], 
	\end{split} \label{eq-expan}
\end{equation}
where $B_0(x)$ and $\psi_0(x)$ are normalized by $B_0$ and $\psi_0$ respectively, and the slowly varying wave amplitudes $ B_{l}^{(n)}$, $v_{xl}^{(n)}$, $v_{yl}^{(n)}$, $v_{zl}^{(n)}$, $\psi_{l}^{(n)}$, etc. satisfy the reality condition: $A_{-l}^{(n)}=A_{l}^{(n)*}$, where the asterisk denotes the complex conjugate, and $k$ and $\omega$ are, respectively, the wave number and the wave frequency of fast carrier MSWs. 
\par
In what follows, we apply the transformations [Eqs. \eqref{eq-stretch} and \eqref{eq-diffop}] and substitute the expansions  [Eq. \eqref{eq-expan}] into the normalized Eqs. \eqref{eq-B1}-\eqref{eq-poiss1}, and then obtain equations for different harmonic modes corresponding to different powers of $\epsilon$ (For some details, see Appendix \ref{appendix}). Here, we assume that the length scales of the density ($L_{\rho_0}$) and magnetic field ($L_{B_0}$)  inhomogeneities are much larger than the length scale $(L)$ of magnetosonic perturbations, i.e., $L_{\rho_0},~L_{B_0}\gg L$. With these assumptions, the imaginary contributions originating from the  inhomogeneities can be ignored.  The results are given in Secs. \ref{sec-sub-DR}-\ref{sec-sub-nls}.
\par
\begin{figure*}
	\centering
	\includegraphics[width=0.8\textwidth]{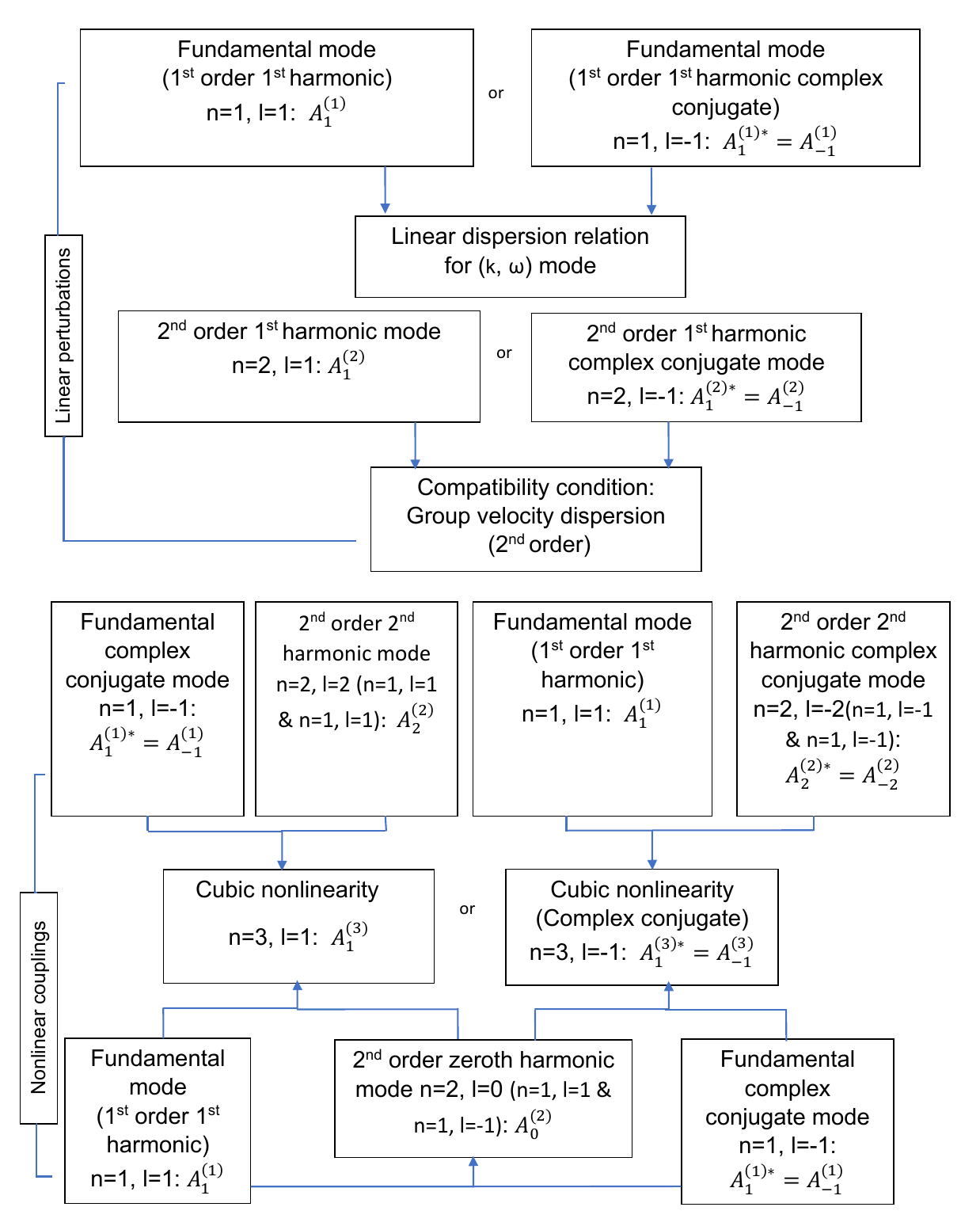}
	\caption{A schematic diagram showing routes to obtain the linear dispersion relation and the group velocity dispersion of the fundamental mode (first-order first harmonic) and the cubic nonlinearity due to nonlinear couplings of the fundamental modes.}
	\label{fig-RPT}
\end{figure*}

 
We use the reductive perturbation technique to derive the NLS equation (with second-order group-velocity dispersion and cubic or Kerr nonlinearity) for the nonlinear evolution of first-order first harmonic (for $n=1,~l=1$) mode or the fundamental mode, i.e., the fast magnetosonic wave. In the linear limit, $l=1$ gives the dispersion relation for the $(\omega, k)$ mode with $n=1$ and the group velocity (second-order) dispersion with $n=2$. However, the second-order zeroth $(l=0)$ harmonic modes can appear due to nonlinear self-interactions of the fundamental modes with complex conjugates ($l=1$ and $l=-1$). In the NLS equation, the cubic nonlinearity for the fundamental mode, i.e., of the form $|A_1^{(1)}|^2 A_1^{(1)}$, appears in the nonlinear interactions between the (i) zeroth harmonic second-order modes $(l=0,~n=2)$ and first harmonic first-order modes or fundamental modes $(l=1,~n=1)$ and (ii) second harmonic second-order modes $(l=2,~n=2)$ and the complex conjugate of the first harmonic first-order modes or fundamental modes $(l=-1,~n=1)$. Thus, the second harmonic second order modes $(l=2,~n=2)$ while interacting with the complex conjugate of the fundamental mode can generate the cubic nonlinearity in the evolution equation for the fundamental mode. 
Although the analysis for the derivation of the NLS equation is standard and well-known, for clarity, we have added a diagram to show routes from linear to nonlinear couplings of different modes (See Fig. \ref{fig-RPT}). 
\subsection{First-order perturbations: Linear dispersion relation} \label{sec-sub-DR}
For the first-order first harmonic perturbations, we obtain from the coefficients of $\epsilon$, the following relations
\begin{equation}
\begin{split}
    &v_{x1}^{(1)}=\frac{\omega }{k} B_1^{(1)},~v_{y1}^{(1)}=-i\frac{ 2}{k} \Omega_{0} \sin\lambda B_1^{(1)},\\
     &v_{z1}^{(1)}=i\frac{2}{k} \Omega_{0} \cos\lambda B_1^{(1)},~\psi_{1}^{(1)}=\frac{{\omega}_J^2 }{k^2} B_1^{(1)},  \label{eq-v1}
          \end{split}
\end{equation}
\begin{equation}
\begin{split}
&-i\omega v_{x1}^{(1)}+ik\left({c}_s^2+{V_A}(x)^2\right)B_1^{(1)}\\
&+2{\Omega}_0\left(\cos\lambda v_{z1}^{(1)}-\sin\lambda v_{y1}^{(1)}\right)-ik\psi^{(1)}=0, \label{eq-vx1}
\end{split}
\end{equation}
where ${V_A}(x)=B_0(x)/\sqrt{\mu_0\rho_0(x)}V_A$ is the normalized inhomogeneous Alfv{\'e}n velocity.
\par 
Next, eliminating the variables and looking for their nonzero solutions, we obtain from Eqs. \eqref{eq-v1}-\eqref{eq-vx1}, the following linear dispersion relation for the fast carrier magnetosonic modes in self-gravitating magnetoplasmas.
\begin{equation}
	\label{eq-disp}
	\begin{split}
	&\omega^2=\left[{c}_s^2+{V_A}^2(x)\right]k^2+4{\Omega}_{0}^2-\omega_J^{2},\\
 \rm{i.e.,}~	&\omega=\left[{c}_s^2+{V_A}^2(x)\right]^{1/2} \left(k^2-k_J^2\right)^{1/2},
	\end{split}
\end{equation}
where $k_J$ is the critical Jeans wave number modified by the Coriolis force, given by,
\begin{equation}
    k_J=\left(\frac{\omega_J^{2}-4 \Omega_{0}^2}{c_s^2 +V_A^2(x)}\right)^{1/2}. \label{eq-kJ}
\end{equation}
From Eq. \eqref{eq-disp}, we note that the fast magnetosonic wave becomes dispersive due to the presence of the term proportional to $\Omega_0^2$, associated with the Coriolis force, and the Jeans instability may occur due to the term proportional to $\omega_J^2$ by the influence of the self-gravitating force  \cite{jeans1902stability}. The instability occurs in the region $k<k_J$,
provided the self-gravity force dominates over the Coriolis force, i.e., $\omega_J>2\Omega_0$. In the other region, i.e., $k>k_J$, the magnetosonic wave can propagate as a real (stable) eigenmode with $\omega_J>2\Omega_0$. On the other hand, in absence of the gravity effects or when the Coriolis force dominates over the self-gravity force with $\omega_J<2\Omega_0$, the MSWs can also propagate as a real mode (without any instability) with the frequency being smaller or larger than the ion-cyclotron frequency and the dispersion relation in the form of high-frequency Langmuir waves in classical plasmas. Thus, it is reasonable to investigate the nonlinear modulation of slowly varying magnetosonic fields by means of a NLS equation.  We also note that the effects of the magnetic field and the density inhomogeneities enter the coefficient of $k^2$ via the Alfv{\'e}n velocity $V_A(x)$, implying that the wave dispersion is greatly modified by the effects of inhomogeneities, and hence the modifications of the phase velocity as well as the group velocity dispersion of magnetosonic envelopes.  
\par
\begin{figure*}
	\centering
	\includegraphics[width=\textwidth]{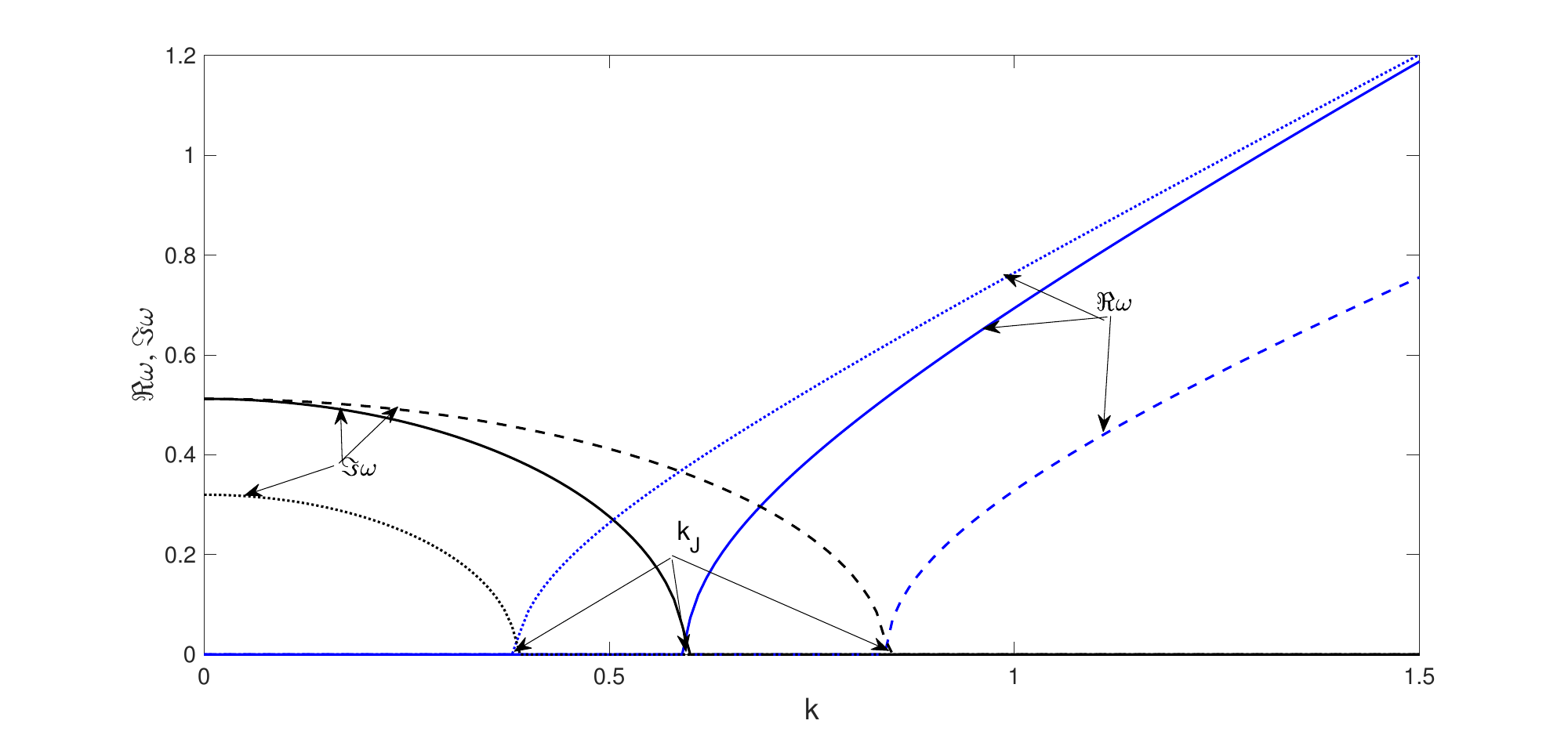}
	\caption{The real $(\Re\omega)$ and imaginary $(\Im\omega)$ parts of the fast magnetosonic wave frequency $\omega$ [Eq. \eqref{eq-disp}] are shown when $\omega_J>2\Omega_0$. The solid, dashed, and dotted lines correspond to the parameter values (i) $\Omega_0=0.8,~\omega_J=2.1\Omega_0,~x=0.5$, (ii) $\Omega_0=0.8,~\omega_J=2.1\Omega_0,~x=1.5$, and (iii) $\Omega_0=0.5,~\omega_J=2.1\Omega_0,~x=1.5$ respectively. The other fixed parameter values are $c_s=0.5$ and $\psi_0=-0.5$.}
	\label{fig1-disp}
\end{figure*}

 
Figure \ref{fig1-disp} displays the profiles of the real wave mode ($\Re\omega$, for $k>k_J$) and the growth rate of instability ($\Im\omega$, for $k<k_J$) by the effects of the density and magnetic field inhomogeneities [See Eq. \eqref{eq-inhom1}], the Coriolis force, and the self-gravity force such that $\omega_J>2\Omega_0$. We find that while the real wave frequency increases with the wave number $k$, the instability growth rate falls off from a nonzero value with $k$ having a cut-off at the critical Jeans wave number, i.e., $k=k_J$. Such a critical value shifts towards a higher value  of $k$ due to the effects of the inhomogeneities in which $\psi_0(x)$ increases, but  $\rho_0(x)$ decreases with increasing values of $x>0$. In this case, the instability domain for $k$ expands and the domain of the real wave mode reduces. Thus, it follows that the density and magnetic field inhomogenities in the background plasma favor the Jeans instability in a wide range of values of $k$, not reported before. On the other hand, a small reduction of the angular frequency $\Omega_0$ of the rotating fluid, associated with the Coriolis force and the Jeans frequency $\omega_J$, can significantly reduce both the growth rate of instability and the instability domain in $k$. We do not consider the case of $\omega_J\approx 2\Omega_0$ at which the fast magnetosonic  wave becomes dispersionless. Such a strict condition may be applicable to low-frequency long-wavelength plasma oscillations. 
\subsection{Second-order perturbations: Compatibility condition} \label{sec-sub-compt}
For the second order ($n=2$) reduced equations with $l=1$, we obtain the following expressions for different harmonic modes 
\begin{equation} 
	\begin{split}
		{v_x}_1^{(2)} &= \frac{\omega}{k} B_1^{(2)}-\frac{i(v_g k -\omega)}{k^2} \frac{\partial B_1^{(1)}}{\partial \xi}, \\
		{v_y}_1^{(2)} &= -\frac{2i\Omega_0 \sin\lambda}{k} B_1^{(2)}+\frac{2 \Omega_0 \sin \lambda}{k^2} \frac{\partial B_1^{(1)}}{\partial \xi},\\
		{v_z}_1^{(2)} &= \frac{2 i \Omega_0 \cos\lambda}{k} B_1^{(2)}-\frac{2 \Omega_0 \cos \lambda}{k^2} \frac{\partial B_1^{(1)}}{\partial \xi},\\
		\psi_1^{(2)} &=  \frac{\omega_{J}^2}{k^2} B_1^{(2)}+ \frac{2i\omega_J^2}{k^3}\frac{\partial B_1^{(1)}}{\partial \xi},
	\end{split}
\end{equation} 
together with the compatibility condition
\begin{equation} 
	\label{eq-vg}
	v_g\equiv\frac{\partial \omega}{\partial k}={\left[c_s^2+V^2_A(x)\right]}/{v_p},
\end{equation}
where $v_p=\omega/k$ is the phase velocity of the carrier magnetosonic waves. From Eq. \eqref{eq-vg}, it is evident that the group velocity $v_g$ of the wave envelope has an inverse relationship with the phase velocity of the carrier wave, i.e., the magnitude of the group velocity increases with a reduction of the magnitude of the phase velocity. Because of the dependency of $v_g$ on the wave frequency $\omega$, the group velocity can be real (imaginary) for $k>k_J~(k<k_J)$ when $\omega_J>2\Omega_0$. It can also be real when $\omega_J<2\Omega_0$ for any real value of $k$. The expression for $v_g$ is clearly modified by the effects of the background magnetic field or density inhomogeneity. Furthermore, having known the characteristics of $\omega$ (See Fig. \ref{fig1-disp}), one can also investigate the features of the real and imaginary parts of $v_g$.    It is seen that the real part of the group velocity decreases with $k>k_J$  and it can be  further reduced (increased) by the effects of the density or magnetic field inhomogeneity (Coriolis force). On the other hand, the imaginary part of $v_g$ is always negative in the domain $k<k_J$ and its magnitude increases with increasing values of $k$.  The detailed analysis of $v_g$ is not necessary at this stage, we will rather focus on the coefficients of the NLS equation to be derived in Sec. \ref{sec-sub-nls}.
\subsection{Third-order perturbations: The NLS equation} \label{sec-sub-nls}
The second-order harmonic modes for $n=2$ and $l=2$ appear  due to the nonlinear self-interaction of the carrier waves, and they are found to be proportional to $[B_1^1]^2$ as
\begin{widetext} 
\begin{equation} 
	\begin{split}
		{v_x}_2^{(2)} &= \frac{\omega}{k} B_2^{(2)}-\frac{\omega}{k} \left[B_1^{(1)}\right]^2, \\
		{v_y}_2^{(2)} &=-\frac{i\Omega_{0} \sin \lambda}{k} \frac{2}{3(4 \Omega_{0}^2-\omega_{J}^{2})}\left[\{2c_s^2+3 V_A^2(x)\}k^2+12\Omega_{0}^2-3\omega_{J}^2\right] \left[B_1^{(1)}\right]^2,\\
		{v_z}_2^{(2)} &=\frac{i\Omega_{0} \cos \lambda}{k} \frac{2}{3(4 \Omega_{0}^2-\omega_{J}^{2})}\left[\{2c_s^2+3 V_A^2(x)\}k^2+12\Omega_{0}^2-3\omega_{J}^2\right] \left[B_1^{(1)}\right]^2,\\
		\psi_2^{(2)} &=\frac{\omega_{J}^2}{4 k^2} \frac{2}{3(4 \Omega_{0}^2-\omega_{J}^{2})}\left[\{2c_s^2+3 V_A^2(x)\}k^2+12\Omega_{0}^2-3\omega_{J}^2\right] \left[B_1^{(1)}\right]^2,\\
		B_2^{(2)} &=\frac{2}{3(4 \Omega_{0}^2-\omega_{J}^{2})}\left[\{2c_s^2+3V_A^2(x)\}k^2+12\Omega_{0}^2-3\omega_{J}^2\right] \left[B_1^{(1)}\right]^2.
	\end{split}
 \end{equation} 
 \end{widetext}
The nonlinear self-interactions of the first-order carrier wave modes also result into the generation of the zeroth harmonic modes. Thus, for $n=2$, $l=0$ we obtain
\begin{equation}
	\begin{split}
		B_0^{(2)}&={v_y}_0^{(2)}={v_z}_0^{(2)}=0,\\
		{v_x}_0^{(2)}&=-\frac{2\omega}{k} |B_1^{(1)}|^2, \\
		\psi_0^{(2)}&=\frac{(2\omega^2+k^2-4\Omega_{0}^2)}{k^2} |B_1^{(1)}|^2.
	\end{split}
\end{equation}
\par
Finally, substituting all the above derived expressions into the third-order harmonic modes ($n=3$ and $l=1$), we obtain the following NLS equation
\begin{eqnarray}
	\label{eq-nls}
	i \frac{\partial B}{\partial \tau} + P \frac{\partial^2 B}{\partial \xi^2} +Q |B|^2 B=0,
\end{eqnarray} 
where we have replaced  $B_1^{(1)}$ by $B$  for simplicity.
The group velocity dispersion coefficient  $P$ and the nonlinear coefficient $Q$, appeared due to the carrier wave self-interactions, are given by 
 \begin{equation}
\label{eq-coeff-P}
\begin{split}
	P\equiv\frac{1}{2}\frac{\partial^2 \omega}{\partial k^2}&=\frac{1}{2} \frac{\left(c_s^2+V_A^2(x)\right)}{\omega^3}\left(4 \Omega_{0}^2-{\omega}_{J}^2\right)\\
	&=\frac{1}{2}\frac{4\Omega_0^2-\omega_J^2}{\left[ c_s^2+V_A^2(x)\right]^{1/2}\left(k^2-k_J^2 \right)^{3/2}},
	\end{split}
\end{equation}
\begin{equation}
\label{eq-coeff-Q}
\begin{split}
	Q=&\frac{1}{2 \omega} \left[\omega^2-\frac{2}{3} \frac{\left(3\omega^2-c_s^2 k^2\right)}{(4 \Omega_{0}^2-{\omega}_{J}^2)}\right.  \\
	&\left. \times  \left\{(2\omega^2+k^2)-\frac{1}{2}\left(4 \Omega_{0}^2-{\omega}_{J}^2\right)\right\}\right]. 
	\end{split} 
\end{equation}   
\par 
From the dispersion Eq. \eqref{eq-disp}, we note that the wave frequency $\omega$ can be either real or purely imaginary depending on whether $k>k_J$ or $k<k_J$. So, the coefficients $P$ and $Q$ can also be either real or purely imaginary. While the coefficient $P$ appears due to the group velocity dispersion and the coupling of the thermal and magnetic pressures with the Coriolis and the Gravity forces, the coefficient $Q$ appears due to the nonlinear self-interactions of higher harmonic modes and coupling between the zeroth- and second-order second harmonic carrier waves. Specifically, the first term of $Q$ appears due to transverse velocity perturbations and the second term of $Q$ is due to the coupling between the zeroth and second harmonic modes. Before proceeding to study the modulational instability and the evolution of magnetosonic wave envelopes, it is pertinent to mention there cases of interest:
\begin{itemize}
\item Case I: $\omega_J<2\Omega_0$, $\forall$ $k$.   
\item Case II: $\omega_J>2\Omega_0$, $k>k_J$.   
\item Case III: $\omega_J>2\Omega_0$, $k<k_J$. 
\end{itemize}  
From the linear analysis in Sec. \ref{sec-sub-DR}, we have noted that while Cases I and II correspond to the Jeans stable mode, Case III corresponds to the Jeans instability. Thus, it is of interest to study the modulational instability conditions in these three cases. Specifically, we will examine whether the Jeans instability region of carrier waves can give rise to the  modulational instability of magnetosonic envelopes.    
\section{MODULATIONAL INSTABILITY} \label{sec-MI}
We follow a similar technique as in \cite{ichikawa1974c} to study the modulation of magnetosonic wave envelopes against a plane wave perturbation. Though the analysis is standard, we reproduce it here for the sake of clarity to the readers. The NLS Eq. \eqref{eq-nls} admits a plane wave time-dependent solution of the form

\begin{equation}
	\label{eq-pwsol}
	B=\eta^{1/2} \exp\left(i\int_a^{\xi} \frac{\sigma}{2P} d\xi\right),
\end{equation}
where $a$ is some constant, and $\eta$ and $\sigma$ are real functions of $\xi$ and $\tau$.
  Substituting Eq. \eqref{eq-pwsol} into Eq. \eqref{eq-nls}, and separating the real and imaginary parts, we obtain
\begin{equation}
	\label{eq-sepate1}
	\frac{\partial \eta }{\partial \tau} +\frac{\partial (\eta \sigma)}{\partial \xi}=0,
\end{equation} 
\begin{equation}
	\label{eq-separte2}
	\begin{split}
		\frac{\partial  \sigma}{\partial \tau} + \sigma\frac{\partial \eta}{\partial \xi}= &2 PQ \frac{\partial \eta}{\partial \xi}  \\
		& +P^2\frac{\partial}{\partial \xi}\left[\eta^{-1/2}\frac{\partial}{\partial \xi}\left(\eta^{-1/2}\frac{\partial \eta}{\partial \xi} \right)\right].
		\end{split}
\end{equation}
Next, we modulate the wave amplitude and phase by small plane-wave perturbations with the wave number $K$ and the wave frequency $\Omega$ as
\begin{equation}
	\label{eq-expansion}
	\begin{split}
		\eta &=\eta_0 + \eta_1 \cos(K\xi-\Omega\tau)+ \eta_2 \sin(K\xi-\Omega\tau),\\
		\sigma &= \sigma_1 \cos(K\xi-\Omega\tau)+ \sigma_2 \sin(K\xi-\Omega\tau), 
	\end{split}
\end{equation}
where $\eta_0$ is a constant, and  $\eta_j$ and $\sigma_j$ for $j=1,2$, are the amplitudes  of perturbations. Thus, the solution \eqref{eq-pwsol} after the modulation according to Eq. \eqref{eq-expansion} can be expressed as 
\begin{equation} \label{eq-Bxt}
\begin{split}
B(x,t)=&\frac{1}{2} \sqrt{\eta_0}\cos(k_0 x-\omega_0 t)\\
&+\chi_1(K)\cos\left[\left(k_0+\epsilon K\right)x-\left(\omega_0+\epsilon v_gK+\epsilon^2\Omega\right)t \right]\\
&+\chi_2(K)\sin\left[\left(k_0+\epsilon K\right)x-\left(\omega_0+\epsilon v_gK+\epsilon^2\Omega\right)t \right]\\
&+\rm{similar~ terms~ with} ~(-K,-\Omega),
\end{split}
\end{equation}
where $\chi_j(K),~j=1,2$ is given by
\begin{equation}
\chi_j(K)=\frac{1}{4}\left(\frac{i}{\sqrt{\eta_0}}+ \frac{\sqrt{\eta_0}}{PK}\sigma_j \right).
\end{equation}
From Eq. \eqref{eq-Bxt}, it is evident that the modulation of magnetosonic wave envelopes is basically a process of three-wave interaction among the unperturbed carrier wave $(k_0,\omega_0)$ and the two side bands with the wave numbers $k_0\pm\epsilon K$ and wave frequencies $\omega_0\pm\epsilon v_g K\pm\epsilon^2\Omega$. Accordingly, the modulational instability (MI) in a nonlinear medium is also known as sideband instability. In general, such instabilities can result from different kinds of nonlinearities in the media. Similar to nonlinear fiber optics, where the MI occurs due to anomalous chromatic dispersion and the Kerr nonlinearity of an optical fiber, the MI in dispersive plasma media is caused by the cubic or Kerr nonlinearity $(Q)$ of an electrostatic or electromagnetic pulse in conjunction with the frequency-dependent group velocity dispersion $(P)$ (i.e., no higher-order dispersion, no propagation loss).  Consequently, the sidebands get amplified in the wave spectrum, leading to increasing oscillations of the wave amplitude. The physical origin of MI of fast magnetosonic waves is the magnetic field-driven ponderomotive force, which can act as a negative pressure to accumulate plasmas in the region where the fast magnetosonic modes exist.
\par 
Substituting the perturbation expansion \eqref{eq-expansion} in Eqs. \eqref{eq-sepate1} and   \eqref{eq-separte2}, and looking for nonzero solutions of $\eta_1$ and $\eta_2$, we obtain the following dispersion relation for the perturbed wave of modulation in self-gravitating rotating magnetoplasmas
\begin{equation}\label{eq-disp-nonl}	
\Omega^2=P^2K^2\left(K^2-\frac{2\eta_0Q}{P}\right).
\end{equation} 
 From the dispersion relation [Eq. \eqref{eq-disp-nonl}] it can be remarked that the frequency becomes imaginary and hence the perturbation grows exponentially (instability) or the small sinusoidal amplitude modulation can be amplified if $PQ>0$ and $K<K_c\equiv\sqrt{2\eta_0Q/P}$ (in the case when $P$ and $Q$ are reals). The gain spectrum can be defined by $\Gamma\equiv\Im\Omega$  so that the power of a wave perturbation grows with time as $\sim \sqrt{\eta} \exp (\Gamma \tau)$.  
In Subsections \ref{sec-sub-caseI}-\ref{sec-sub-caseIII}, we will study the conditions for the modulational instability and the instability growth rate in the three different cases as mentioned above.  
\subsection{Case I:  $\omega_J<2\Omega_0$, $\forall$ $k$} \label{sec-sub-caseI}
We consider the case when the contribution from the Coriolis force dominates over the self-gravitating force, i.e. when $\omega_J<2\Omega_0$. In this case, the carrier Jeans wave frequency  $\omega$ becomes real for any values of the wave number $k$ [See Eq. \eqref{eq-disp}], and so are $P$ and $Q$. Also, from Eq. \eqref{eq-coeff-P}, it is evident that that  $P>0$ for  $\omega_J<2\Omega_0$.  Furthermore, inspecting on the expression of $Q$ [Eq. \eqref{eq-coeff-Q}], we find that when $\omega_J<2\Omega_0$ holds, the second term of $Q$ becomes positive and larger than the first term, giving $Q<0$. Thus, in this case, $PQ<0$, and Eq. \eqref{eq-disp-nonl} gives a real wave frequency $\Omega$ for the perturbation of modulation, implying that the slowly varying magnetosonic wave envelope is always stable under the modulation. It follows that the modulated magnetosonic wave is also stable in absence of the self-gravity force.    
\subsection{Case II:  $\omega_J>2\Omega_0$, $k>k_J$}
\label{sec-sub-caseII}
\begin{figure*}
	\centering
	\includegraphics[width=\textwidth]{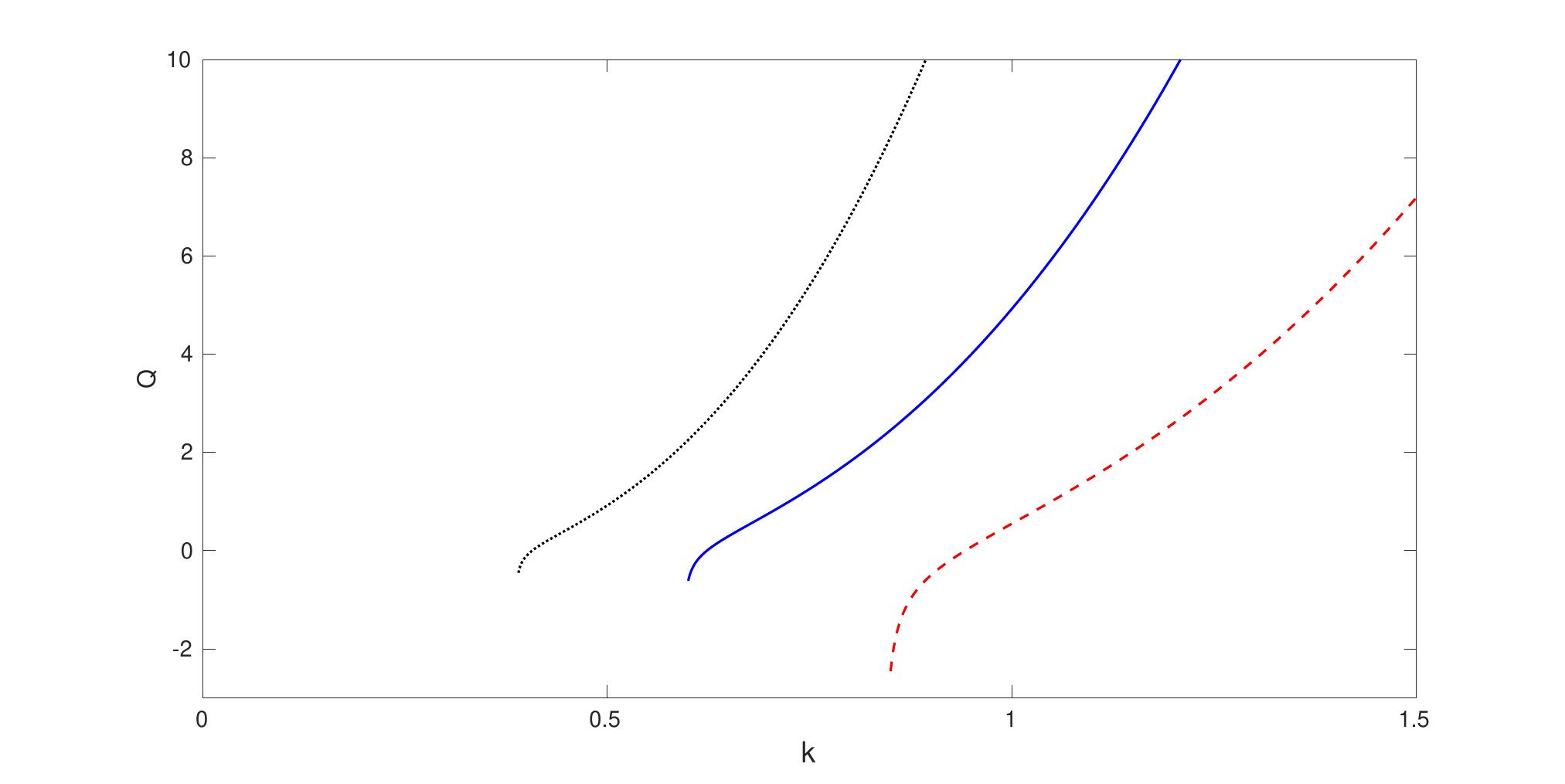}
	\caption{Plot of $Q$ [Eq. \eqref{eq-coeff-Q}] vs $k$ is shown for Case II: $\omega_J>2\Omega_0$, $k>k_J$. The fixed parameter values and different parameter values for the solid, dashed, and dotted lines are the same as for Fig. \ref{fig1-disp}. Since $P<0$ $\forall k$, the stable and unstable regions in $k$ are corresponding to $Q>0$ and $Q<0$ respectively.  }
	\label{fig4-Q}
\end{figure*}

We consider the case when the self-gravitating force dominates over the Coriolis force, i.e. $\omega_J>2\Omega_{0}$ and $k>k_J$. In this case, as discussed before in Sec. \ref{sec-sub-DR}, since the carrier wave frequency $\omega$ is real, the coefficients $P$ and $Q$ of the NLS equation are also real. However, in contrast to Case I, $P<0$ and $Q$ can be both positive and negative for $k~(>k_J)$, so that one can have both stable $(PQ<0)$ and unstable $(PQ>0)$ regions. Figure \ref{fig4-Q} shows the plot of $Q$ versus the carrier wave number $k$. We find that due to the inhomogenity effects, the domain of $k$ in which $Q<0$ (for which $PQ>0$ and the modulational instability occurs) expands and the domain shifts towards lower values of $k$ due to a reduction of the magnitude of $\Omega_0$. For example, for  
(i) $\Omega_0=0.8,~\omega_J=2.1\Omega_0,~x=0.5$, we have $Q_1<0~(>0)$ in $0.6\lesssim k<0.624~(k\gtrsim0.624)$ (See the solid line of Fig. \ref{fig4-Q}) (ii) $\Omega_0=0.8,~\omega_J=2.1\Omega_0,~x=1.5$, we have  $Q_1<0~(>0)$ in $0.85\lesssim k<0.94~(k\gtrsim0.94)$ (See the dashed line of Fig. \ref{fig4-Q}), and (iii) $\Omega_0=0.5,~\omega_J=2.1\Omega_0,~x=1.5$, we have $Q_1<0~(>0)$ in $0.39\lesssim k<0.41~(k\gtrsim0.41)$ (See the dotted line of Fig. \ref{fig4-Q}). The other fixed parameter values are $c_s=0.5$ and $\psi_0=-0.5$. 
\par 
In the case of $PQ>0$ for which the modulational instability occurs, the instability growth rate $(\Gamma)$ can be obtained from Eq. \eqref{eq-disp-nonl} for $K<K_c$ as 
\begin{equation}\label{eq-growth}	
\Gamma=|P|K\sqrt{K_c^2-K^2},
\end{equation} 
where $K_c=\sqrt{2\eta_0Q/P}$ is the critical wave number of perturbation, and the maximum growth rate, $\Gamma_{\max}=\eta_0|Q|$  is attained at $K=K_c/\sqrt{2}$. 
\par
\begin{figure*}
	\centering
	\includegraphics[width=\textwidth]{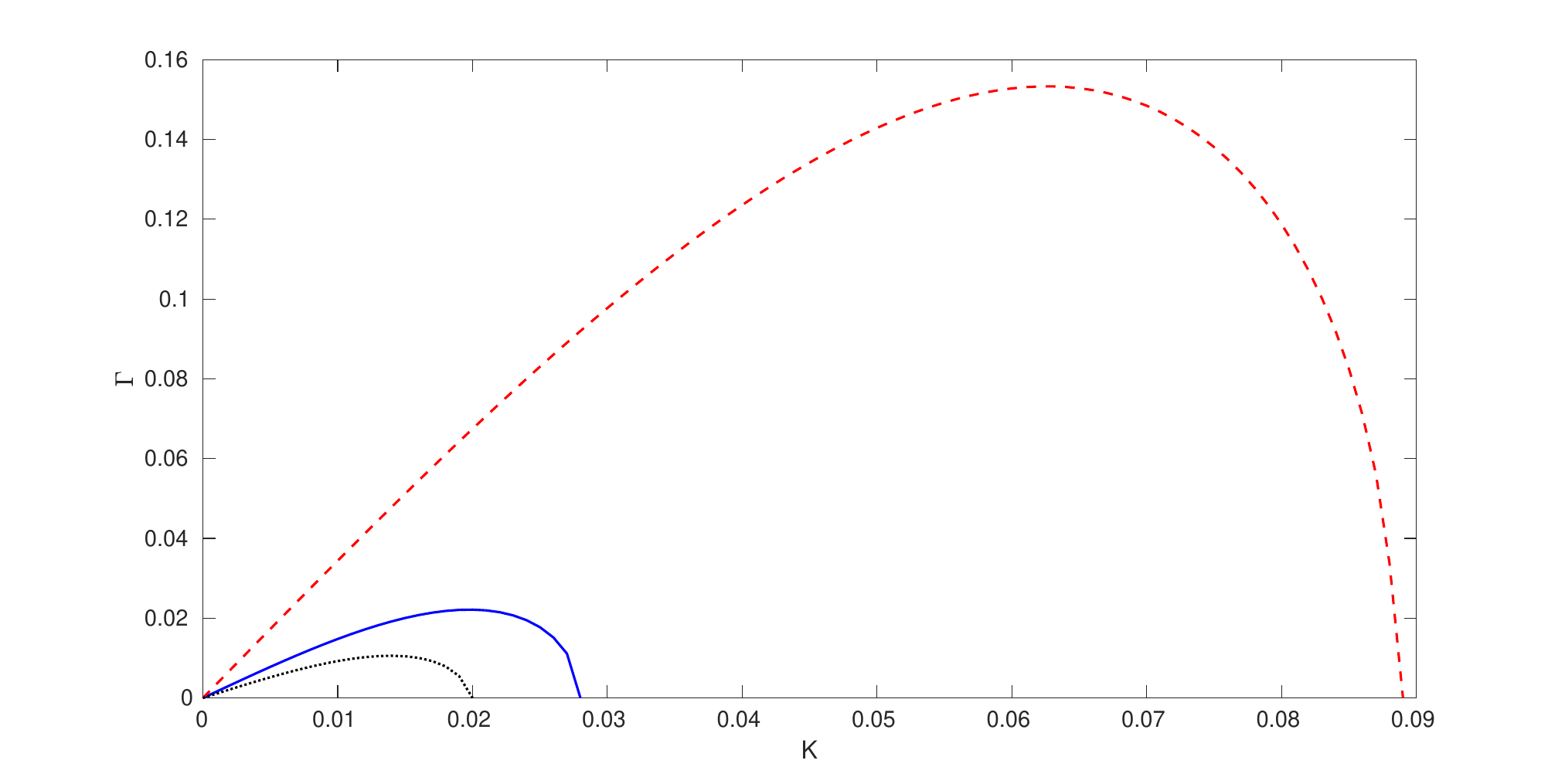}
	\caption{The modulational instability $(PQ>0)$ growth rate $\Gamma$ [Eq. \eqref{eq-growth}] is shown against the  modulation wave number $K$ of perturbation.  The fixed parameter values and different parameter values for the solid, dashed, and dotted lines are the same as for Fig. \ref{fig1-disp}.  }
	\label{fig5-growth}
\end{figure*}

 
Figure \ref{fig5-growth} displays the growth rate of instability [Eq. \eqref{eq-growth}] for different parameter values as in Fig. \ref{fig1-disp}. We find that the instability growth rate can be enhanced and maximized by the effects of the magnetic field or density inhomogeneity. However, it can be reduced or minimized by reducing the angular frequency $\Omega_0$ and the Jeans frequency.
\subsection{Case III:  $\omega_J>2\Omega_0$, $k<k_J$}
\label{sec-sub-caseIII}
Here, we consider the most interesting case when $\omega_J>2\Omega_0$, but $k<k_J $. From the analysis in Sec. \ref{sec-sub-DR}, it is evident that the carrier wave frequency $\omega$ is purely imaginary (Jeans instability) and so are the coefficients $P$ and $Q$ of the NLS equation. Thus, assuming $P=iP_1$ and $Q=i Q_1$,  Eq. \eqref{eq-disp-nonl} gives
\begin{equation}
	\begin{split}
		\label{eq-disp-nonl2}	
		\Omega= P_1 K\sqrt{\frac{2\eta_0 Q_1}{P_1}-K^2},
	\end{split}
\end{equation} 
where $P_1$ and $Q_1$ are reals, given by,
 \begin{equation}
\label{eq-P1-Q1}
\begin{split}
	&P_1=\frac{1}{2}\frac{\omega_J^2-4\Omega_0^2}{\left[ c_s^2+V_A^2(x)\right]^{1/2}\left(k_J^2-k^2 \right)^{3/2}},\\
	&Q_1=-\frac{1}{2 \gamma} \left[-\gamma^2+\frac{1}{3} \frac{3\gamma^2+c_s^2 k^2}{\omega_J^2-4 \Omega_{0}^2} \right.\\
	&  \times \left. \left(4\gamma^2-2k^2-\left(\omega_J^2-4 \Omega_{0}^2\right)\right)\right], 
	\end{split}
\end{equation}   
with 
\begin{equation}
\begin{split}
\gamma\equiv\Im\omega&=\sqrt{\left[c_s^2+V_A^2(x)\right]\left(k_J^2-k^2\right)}\\
&=\sqrt{\left[c_s^2+V_A^2(x)\right]k^2+\omega_J^2-4\Omega_0^2}.
\end{split}
\end{equation}
\par
\begin{figure*}
	\centering
	\includegraphics[width=\textwidth]{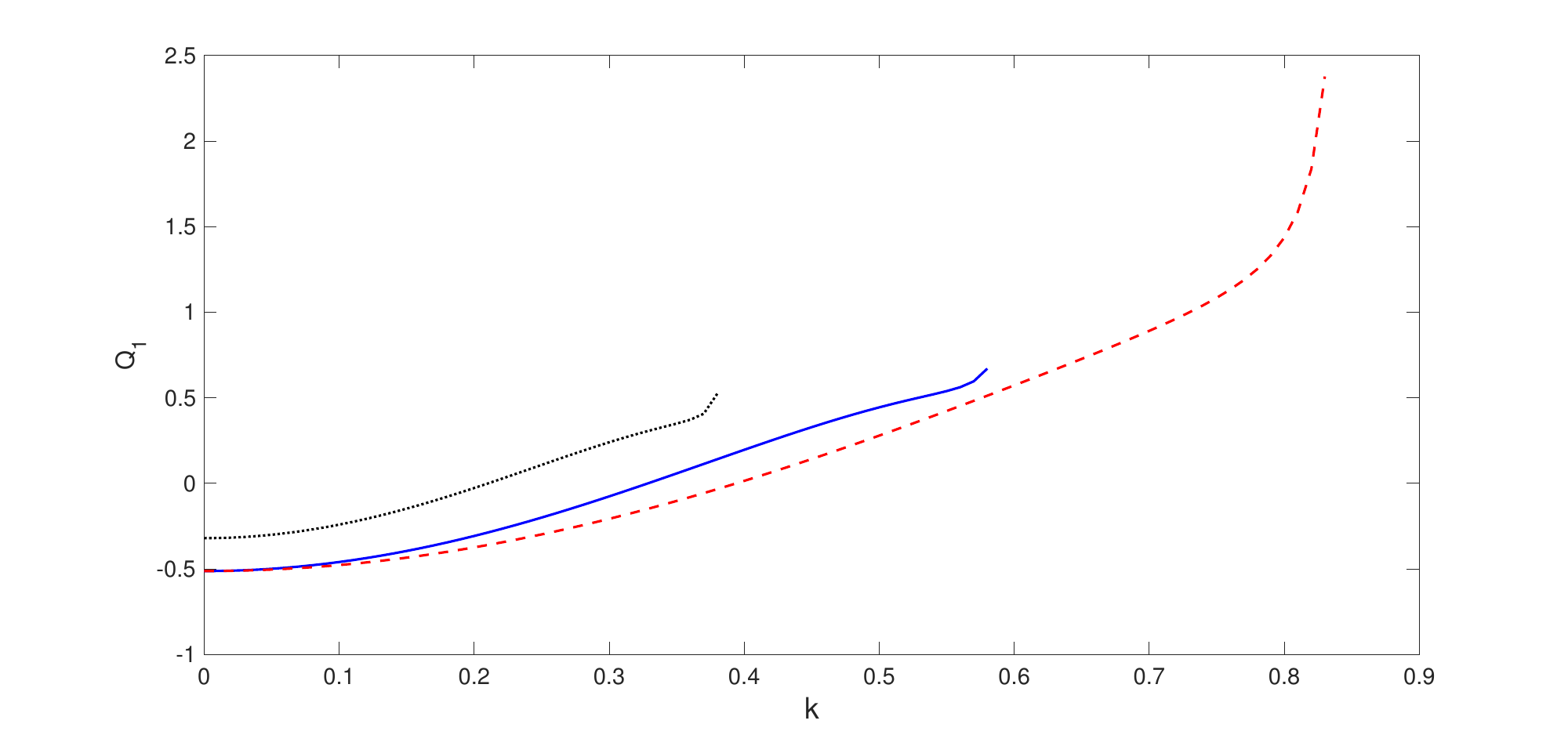}
	\caption{ Plot of $Q_1$ versus $k$ is shown for Case III:  $\omega_J>2\Omega_0$, $k<k_J$. The fixed parameter values and different parameter values for the solid, dashed, and dotted lines are the same as for Fig. \ref{fig1-disp}. }
	\label{fig2-Q1}
\end{figure*}

 
 From Eq. \eqref{eq-disp-nonl2}, we find that the instability condition depends on the sign of the product $P_1Q_1$ and/or on the value of $K$ smaller or larger than a critical value $K_c$.
  When $P_1Q_1<0$, Eq. \eqref{eq-disp-nonl2} gives $\Omega$ purely imaginary for all values of $K$ and so we have the modulational instability for all $K$, but with $k<k_J$. The corresponding growth rate of instability can be obtained by setting $\Omega=i\Gamma_1$ as 
 \begin{equation}
 \label{eq-Gam1}
	\Gamma_1=\big|P_1\big| K\sqrt{2\eta_0 \Big|\frac{Q_1}{P_1}\Big|+K^2}.
\end{equation}
On the other hand, when $P_1Q_1>0$, the modulational instability occurs for $K>K_c$, where $K_c$ is the critical wave number, given by,
\begin{equation}\label{eq-Kc}
    K_c=\sqrt{2\eta_0 \Big|\frac{Q_1}{P_1}\Big|}.
\end{equation}
The corresponding instability growth rate is given by 
\begin{equation} \label{eq-Gam2}
	\Gamma_2=|P_1| K\sqrt{K^2-2\eta_0\Big|\frac{Q_1}{P_1}\Big|}.
\end{equation} 
However, the modulated wave is stable for $K<K_c$. The instability condition stated above is in contrast to the typical condition ($PQ>0$, $K<K_c$) of modulational instability of wave envelopes in plasmas without self-gravity effects.  Inspecting on the coefficients $P~(=iP_1)$ and $Q~(=iQ_1)$, we find that  $P_1$ is always positive for $k<k_J$ and $\omega_J>2\Omega_0$. From Fig. \ref{fig2-Q1}, we find that $Q_1$ can be both positive and negative in a finite domain of $k~(<k_J)$. The domain for $Q_1>0$ in which the modulational instability occurs expands due to the effects of magnetic field or density inhomogeneity and shrinks due to reduction of the contribution from the Coriolis force compared to the self-gravitating force. For example, for  
(i) $\Omega_0=0.8,~\omega_J=2.1\Omega_0,~x=0.5$, we have $Q_1<0~(>0)$ in $0<k<0.33~(0.33\lesssim k\lesssim0.594)$ (See the solid line of Fig. \ref{fig2-Q1}) (ii) $\Omega_0=0.8,~\omega_J=2.1\Omega_0,~x=1.5$, we have  $Q_1<0~(>0)$ in $0<k<0.4~(0.4\lesssim k\lesssim0.842)$ (See the dashed line of Fig. \ref{fig2-Q1}), and (iii) $\Omega_0=0.5,~\omega_J=2.1\Omega_0,~x=1.5$, we have $Q_1<0~(>0)$ in $0<k<0.22~(0.22\lesssim k\lesssim0.386)$ (See the dotted line of Fig. \ref{fig2-Q1}). The other fixed parameter values are $c_s=0.5$ and $\psi_0=-0.5$.

\begin{figure*}
	\centering
	\includegraphics[width=\textwidth]{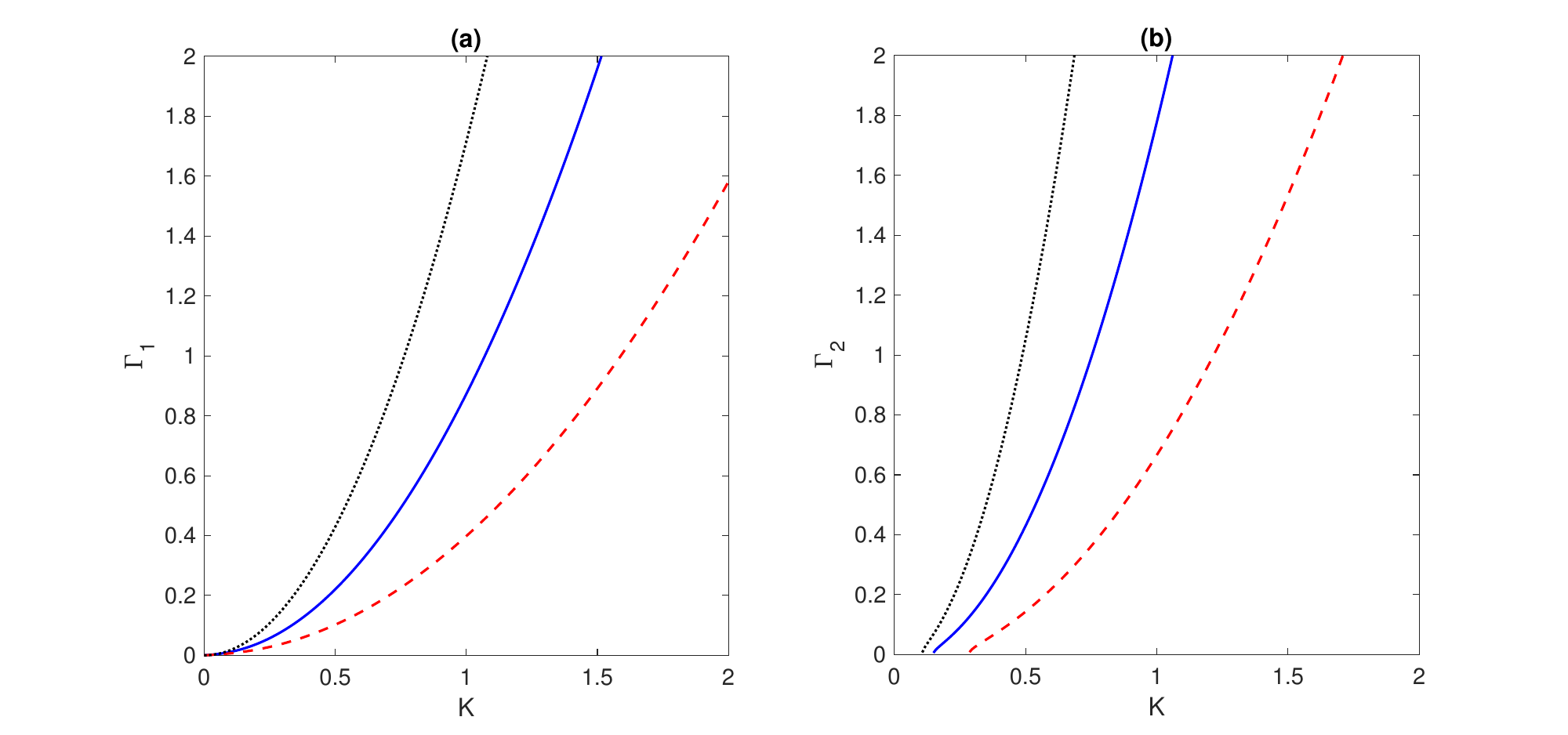}
	\caption{The instability growth rates (a) $\Gamma_1$ [Eq. \eqref{eq-Gam1}, when $P_1Q_1<0$] and (b) $\Gamma_2$ [Eq. \eqref{eq-Gam2}, when $P_1Q_1>0$] are plotted against the modulation wave number $K$ of perturbation.  The fixed parameter values and different parameter values  the solid, dashed, and dotted lines are the same as for Fig. \ref{fig1-disp}. }
	\label{fig3-Gam12}
\end{figure*}
\par 
Figure \ref{fig3-Gam12} displays the growth rates  $\Gamma_1$ and $\Gamma_2$ [Eqs. \eqref{eq-Gam1} and \eqref{eq-Gam2}] corresponding to $P_1Q_1<0$ and $P_2Q_2>0$ respectively. Evidently, we have a growing modulational instability without any cut-off at any value of $K$. The growth rate becomes higher the larger is the Jeans frequency compared to the angular velocity. 
\section{Applications and discussions} \label{sec-appl}
The solar corona  is typically a complex system in the self-gravitating field of the Sun, and has been known to be an active medium for the dynamics and stability of magnetosonic waves, since one of the most important problems of coronal heating is associated with the wave \cite{kolotkov2021}.  Recent Solar Optical Telescope (SOT) observations of small-scale oscillations also indicate the existence of fast magnetosonic waves in solar prominence threads and pillars \cite{ofman2023}. Past observations in the X-ray corona also reported slowly moving perturbations  that can propagate with the speed $400$ km/s during the first nine minutes of filament disruption \cite{rust1979}. Such a velocity seems to be close to the Alfv{\'e}n speed of fast magnetosonic waves in a low-$\beta$ plasma. The speed was, however, reduced to $190$ km/s and $20$ km/s after twenty minutes and four hours respectively. To explain this observation, one can assume that the fast magnetosonic waves originating from disrupted filaments can propagate  as slowly moving magnetosonic wave envelopes through the modulational instability \cite{sakai1983modulational}.
\par 
 Typical solar coronal plasma parameters are \cite{kolotkov2021} (i) the number density of electrons/ions, $n\sim10^9~\rm{cm}^{-3}$, (ii) the electron temperature, $T_e\sim10^{6}$ K, and (iii)  the plasma $\beta\sim0.2~(\beta\rightarrow0)$ at the magnetic field strength, $B_0\sim 4$ G  ~($40$ G). 
For illustration purpose, we calculate the maximum growth rate in the case of $\omega_J>2\Omega_0$, $k>k_J$ (Case II in Sec. \ref{sec-sub-caseII}) at $B_0=4$ G as $\Gamma_{\rm{max}}\sim0.0076$  (for $\eta_0=0.01$) such that the typical growth time can be estimated as $\tau_{\rm{growth}}\sim4$ s, i.e., faster than predicted before \cite{sakai1983modulational} and may be reasonable to explain the transition time (twenty minutes) from fast magnetosonic mode to slowly varying envelopes. The typical Jeans wavelength and the Alfv{\'e}n velocity can be estimated as   $\lambda_J\sim2\times10^4$ m and  $V_A\sim3\times10^5$ m/s.  We can also estimate the group velocity as $v_g\sim 5\times10^5$ m/s. These velocities are close to the observational velocity $190$ km/s of slowly moving perturbations. The instability growth rate can be significantly high by the effects of the magnetic field and density inhomogeneities where the Alfv{\'e}n velocity $V_A(x)$  gets reduced.  Note that the above results are valid for low-$\beta$ plasmas where the length scale of magnetic field inhomogeneity is much smaller than that of the density inhomogeneity and the wavelength associated with the fast carrier magnetosonic mode is small compared to the wavelength of the slow modulation along the magnetic field.   
On the other hand, the typical growth rate can be high having no cut-off at any wavelength of modulation, and the corresponding time scale can be low in the case when $\omega_J>2\Omega_0$ and $k<k_J$ (Case III in Sec. \ref{sec-sub-caseII}). However, to be consistent with the small perturbations, the admissible growth rate should be at a wavelength  (of slow modulation) above its critical value.
\par 
It has been established that the fast magnetosonic mode can act as a candidate for coronal loop heating through the exchange of energy and momentum \cite{kolotkov2021}. The present theory of modulational instability should help predict the energy transfer rate of slowly varying magnetosonic envelopes in solar plasmas under the relative influences of the self-gravity force and the  Coriolis force and the excitation of fast magnetosonic carrier modes at length scales below or above the Jeans critical length.  
\par
\begin{figure*}
	\centering
	\includegraphics[width=\textwidth]{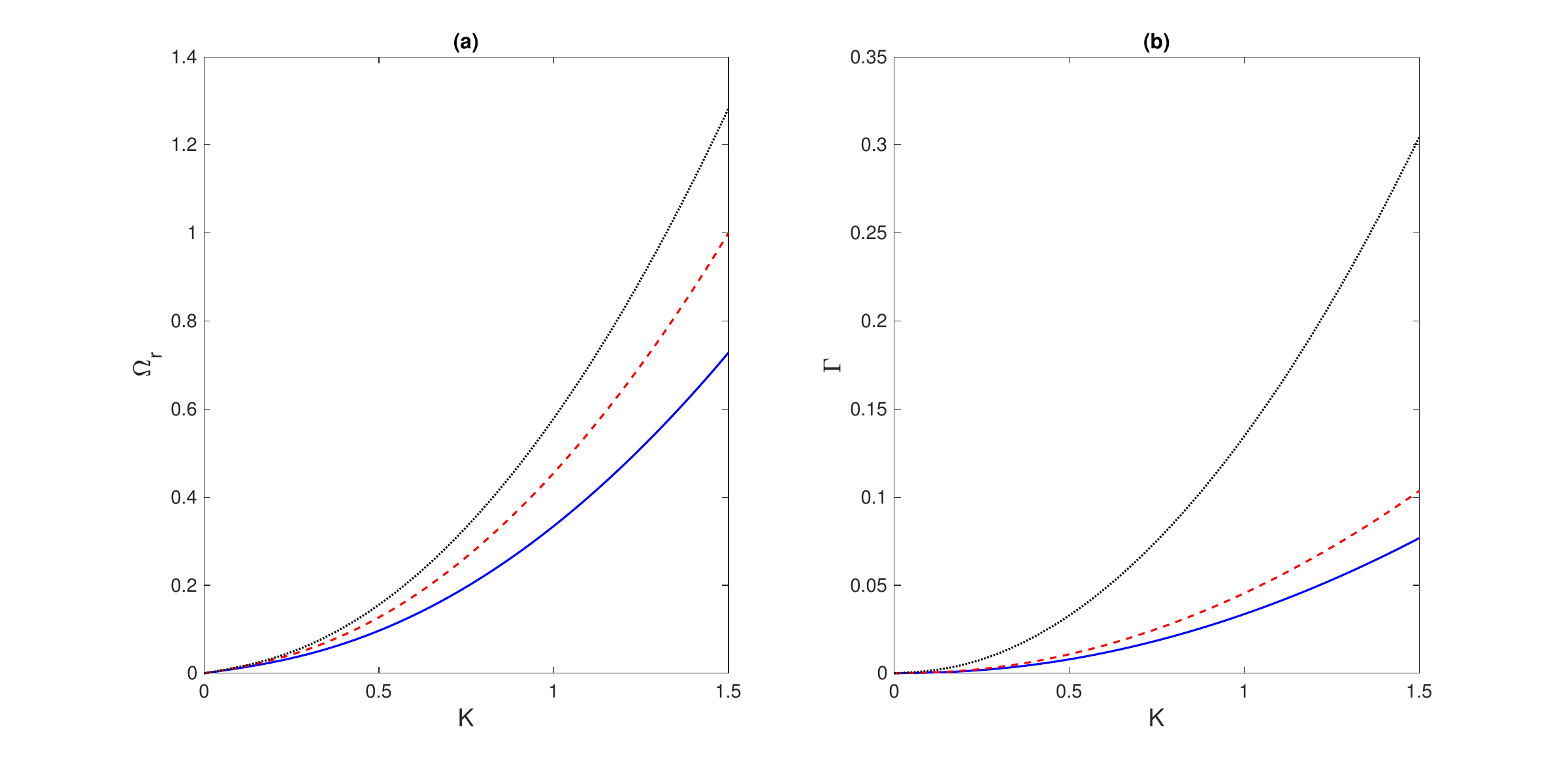}
	\caption{The frequency shift [subplot (a), Eq. \eqref{eq-freq-shift}] and the instability growth rate  [subplot (b), Eq. \eqref{eq-growth-g}] are plotted against the modulation wave number $K$ of perturbation.  The solid, dashed, and dotted lines correspond to the parameter values as for Fig. \ref{fig1-disp}. The other fixed parameter (normalized) values are $g=0.5,~c_s=0.5$, and $k=0.4$.}
	\label{fig6-g}
\end{figure*}

 
While the self-gravity force works, e.g., in the inner layers of the Sun: The core, the radiative zone, and the convective zone, or the cores of giant stars, the constant gravity may apply to regions of planetary atmosphere (e.g., Earth’s atmosphere) or solar atmosphere [e.g., Sun’s atmosphere (outer): Photosphere,  Chromosphere, and Corona].  
 Thus, in the case of plasmas with constant gravity, the gravity force, $\mathbf{g}=-\nabla\psi$ is to be replaced by $\mathbf{g}=(-g,0,0)$ and there will be no gravitational Poisson equation. Even if it is considered for the sake of clarity, it will result into the quasineutrality condition, already assumed for the derivation of the MHD equations. In this case, the term $-\omega_J^2$ in the dispersion equation \eqref{eq-disp} for the fast magnetosonic carrier waves  will be replaced by $igk$, which will result into the real wave frequency, given by, 

 \begin{equation}
	\label{eq-disp-g}
	 \omega_r^2=\left[{c}_s^2+{V_A}^2(x)\right]k^2+4{\Omega}_{0}^2  
	\end{equation}
and the instability growth rate, $\gamma=gk/\omega_r$. From the inhomogeneous equilibrium state one can obtain by the same assumption as for the fluid density in Sec. \ref{sec-sub-eqbm} as $\rho_0(x)\approx \rho_0+gx$.  The coefficients of the group velocity dispersion $(P)$ and the nonlinearity $(Q)$ in the NLS equation will be complex, given by,
\begin{equation} \label{eq-P-g}
P=\frac{1}{2\omega^2}\left[\left({c}_s^2+{V_A}^2(x)\right)(\omega-v_g k)-\frac12igv_g \right],
\end{equation} 
\begin{equation}\label{eq-Q-g}
Q=\frac{1}{2\omega}\left(\omega^2-\frac{\omega_1\omega_2}{6\Omega_0^2+igk}\right),
\end{equation}
where $\omega_1=\left(3 V_A^2(x)+2c_s^2\right)k^2+12\Omega_0^2+2igk$ and $\omega_2=\left(3 V_A^2(x)+2c_s^2\right)k^2+6\Omega_0^2+2igk$. Consequently, instead of the purely growing instability as in Case III of Sec. \ref{sec-sub-caseIII}, we have the frequency up-shift $(\Omega_r)$ and the modulational instability growth rate $(\Gamma)$, given by,
\begin{equation}\label{eq-freq-shift}
	\Omega_r=\frac{1}{\sqrt{2}}\left[ -N + \sqrt{N^2+4M^2}\right]^{1/2},
\end{equation}
\begin{equation} \label{eq-growth-g}
	\Gamma=\frac{1}{\sqrt{2}} \left[ N +\sqrt{N^2+4M^2}\right]^{1/2},
\end{equation}
where the expressions for $M$ and $N$ are 
\begin{equation}
	M=K^2\left\{ P_1 P_2 K^2-\eta_0 (P_2 Q_1+Q_2 P_1)\right\},
\end{equation}
\begin{equation}
	N=K^2\left\{2 \eta_0 (P_1 Q_1+P_2 Q_2)- (P_1^2-P_2^2) K^2\right\}.
\end{equation}
For brevity, we have presented the characteristics of $\Omega_r$ and $\Gamma$ as shown in Fig. \ref{fig6-g}. It is found that both assume higher values at a reduced angular frequency $\Omega_0$ (the dotted lines) and by the effects of the density inhomogeneity (the dashed lines). Also, the growth rate $\Gamma$ is similar to Case III of Sec. \ref{sec-sub-caseIII} in self-gravitating plasmas, i.e., it grows with the wave number of modulation $K$ without any cut-offs.
\par 
Thus, the instabilities of fast magnetosonic modes and the MI of solwly varying wave envelopes under the influences of self-gravity force and the constant gravity force are quite distinctive. While the Jeans critical wavelength appears above which the Jeans instability of fast carrier modes occurs and hence the MI of magnetosonic envelopes in self-gravitating fields, the fast modes appears to be unstable by the constant gravity force without any restriction of the wavelength unless the length at which the collective behaviors disappear. Furthermore, in contrast to plasmas under constant gravity, the stable Jeans mode in self-gravitating fields may give rise to the MI of slowly varying envelopes with a finite growth rate  and cut-off at a finite wave number of modulation (See Case II in Sec. \ref{sec-sub-caseII}).  
\section{Summary and conclusion} \label{sec-summary}
To summarize, we have studied the dispersion properties of fast magnetosonic waves and the modulation of slowly varying magnetosonic wave envelopes in inhomogeneous magnetoplasmas under the relative influence of the gravitational force and the Coriolis force due to rotating fluids. Using the multiple-scale reductive perturbation technique, we have derived the NLS equation for the evolution of slowly varying magnetosonic envelopes. It is found that the inhomogeneities in the background magnetic field and the fluid density favor the Jeans instability to occur in a larger domain of the carrier wave number $(k)$ below the Jeans critical wave number $(k_J)$. The latter is, however, defined when the self-gravity force dominates over the Coriolis force or the  Jean frequency $\omega_J$ becomes larger than the increased angular frequency $2\Omega_0$.  On the other hand, when the Coriolis force dominates over the self-gravity force or is dominated by the self-gravity force with $k>k_J$, the fast magnetosonic carrier mode is always stable. It is interesting to note that the MSW envelope, corresponding to the unstable carrier Jeans mode with $\omega_J>2\Omega_0$ and $k<k_J$, is always unstable under the plane wave modulation having no cut-offs in the growth rates.  However, the stable Jeans mode with $\omega_J>2\Omega_0$ but $k>k_J$ can lead to modulational stability or instability with finite growth rates and cut-offs. Such an instability growth rate is enhanced by the effects of the magnetic field or density inhomogeneities.
It is important to note that in self-gravitating rotating plasmas, when the Coriolis force dominates over the self-gravity force ($\omega_J<2\Omega_0$), the carrier Jeans magnetosonic mode is always stable and so is the modulation perturbation of magnetosonic envelopes. In this case, the Coriolis force plays a role in stabilizing both the linear (Jeans) and nonlinear (modulational) instabilities. On the other hand, when the Coriolis force is dominated by the gravity force ($\omega_J>2\Omega_0$ and $k<k_J$), the Coriolis force itself or both the forces can not stabilize the instabilities but can reduce the corresponding growth rates of instabilities either at higher values of $\Omega_0$ or at lower values of both $\omega_J$ and $\Omega_0$.   However, both the forces can stabilize the modulational instability at lower values of both $\omega_J$ and $\Omega_0$ even when  $\omega_J>2\Omega_0$ but $k>k_J$ (the case where the carrier Jeans mode is linearly stable).     
 We have also discussed the possible applications of modulational instability in solar plasmas, such as those in the X-ray corona and solar prominence, and noted that the estimated Alfv{\'e}n velocity of fast modes and the group velocity of slowly varying envelopes are close to the observational values. Also, the estimated growth time is found to be reasonable to explain the transition time from the fast magnetosonic modes to slowly varying envelopes at the nonlinear stage.
\par 
As an illustration, we have also examined the influence of the constant gravity force instead of the self-gravitation. It is found that the fast carrier magnetosonic mode is always unstable, leading to a frequency up-shift of the frequency of modulation (instead of a purely unstable mode) and the instability growth rate having no cut-offs, similar to Case III of self-gravitating plasmas.
\par 
Modulational instability is a prerequisite for energy localization and the structure formation of different types of envelope solitons in the NLS equation. The unstable and stable modulation domains reported here could be helpful to look for bright and dark envelope solitons, Kuznetsov–Ma breather soliton, Akhmediev breather, as well as Peregrine solitons \cite{roy2023}. We have limited our discussion to the one-dimensional propagation of magnetosonic waves and the modulation along the carrier wave vector. However, considering the arbitrary direction of propagation and the magnetic field in a plane can give rise to Alfvén and magnetosonic modes as fundamental modes that can get coupled via the Coriolis force for rotating fluids \cite{turi2022magnetohydrodynamic} or due to Hall effects in Hall MHD plasmas \cite{ruderman2020}. In this case, any one (to be modified by the other) can be a fundamental mode to study the modulational instability of wave envelopes. The amplitude modulation obliquely to the wave vector could result in different instability criteria with stable and unstable regions \cite{misra2011} that would differ from those presented here. Also, in multi-dimensional propagation of magnetosonic waves, the modulation along a wave vector may not be described by a single NLS-type equation but a coupled set of equations, i.e.,  Davey–Stewartson-like equations \cite{misra2011}. The latter can govern the possible collapse of magnetosonic wave envelopes. Thus, oblique modulation of multi-dimensional magnetosonic wave envelopes in Hall MHD rotating plasmas could be a project for future studies. 
\par 
To conclude, since the fast magnetosonic modes can be a candidate for coronal loop heating in which the momentum and energy transfer occur, the theoretical results should help predict the growth time of slowly varying magnetosonic envelopes that can be transformed from the fast magnetosonic modes emanating from disrupted filaments in the X-ray corona through nonlinear interactions. The higher the growth rate shorter is the time scale for the instability. It is to be noted that the proposed coronal heating mechanism has been the collisionless Landau damping. The fast magnetosonic modes propagating parallel or perpendicular to the magnetic field cannot suffer collisionless damping, but may be damped due to viscosity effects \cite{sakai1983modulational}. However, it has been shown that modulational instability is the dominant mechanism for energy transfer in the coronal loops than the viscous damping \cite{sakai1983modulational}. The detailed investigation in this direction is beyond the scope of the present study. Accordingly, multi-dimensional propagation of magnetosonic waves in dissipative plasmas could be a project of future study. 

\section*{Acknowledgments}
One of us, J. Turi, wishes to thank the Council of Scientific and Industrial Research (CSIR) for a Senior Research Fellowship (SRF) with reference number 09/202(0115)/2020-EMR-I. The authors also thank Sima Roy of the University of Engineering \& Management, Kolkata, for rechecking the derivations of different expressions for the NLS equation and the anonymous referees for some helpful comments, which improved the manuscript in its present form.
\section*{Author declarations}
\subsection*{Conflict of Interest}
The authors have no conflicts to disclose.
\subsection*{Author Contributions}
\textbf{J. Turi:} Writing--draft, Methodology, Investigation, Formal analysis. \textbf{A. P. Misra:} Writing--draft, review \& editing, Validation, Methodology, Investigation, Formal analysis, Conceptualization. 
\section*{Data availability statement}
All data that support the findings of this study are included within the article (and any supplementary files).
\appendix
\section*{n-th order reduced equations} \label{appendix}
We apply the transformations [Eqs. \eqref{eq-stretch} and \eqref{eq-diffop}] and substitute the expansions  [Eq. \eqref{eq-expan}] into Eqs. \eqref{eq-B1}-\eqref{eq-poiss1}, to get the following set of  reduced n-th order equations:
\begin{widetext} 
	\begin{equation}
		\begin{split}
			-i\omega l B_{l}^{(n)}-\lambda \frac{\partial}{\partial \xi} B_{l}^{(n-1)}+\frac{\partial}{\partial \tau} B_{l}^{(n-2)} +ilk\left[B_0(x) {v_x}_{l}^{(n)} + \sum_{n'=1}^{\infty} \sum_{l'=-\infty}^{+\infty}{v_x}_{l-l'}^{(n-n')} B_{l'}^{(n')}\right]\\
			+\frac{\partial B_0(x) }{\partial x} v_{xl}^{(n)} + B_0(x) \frac{\partial}{\partial \xi} v_{xl}^{(n-1)} + \sum_{n'=1}^{\infty} \sum_{l'=-\infty}^{+\infty}\frac{\partial}{\partial \xi} ({v_x}_{l-l'}^{(n-n')} B_{l'}^{(n'-1)} )=0\\ 
		\end{split}
	\end{equation}
\begin{equation}
	\begin{split}
		-i\omega l B_0(x) v_{xl}^{(n)}-\lambda B_0(x) \frac{\partial}{\partial \xi} v_{xl}^{(n-1)}+ B_0(x) \frac{\partial}{\partial \tau} v_{xl}^{(n-2)}+\sum_{n'=1}^{\infty} \sum_{l'=-\infty}^{+\infty}\left[ -i\omega l' {B}_{l-l'}^{(n-n')} {v_x}_{l'}^{(n')} 
	-\lambda {B}_{l-l'}^{(n-n')} \frac{\partial}{\partial \xi} {v_x}_{l'}^{(n'-1)}\right.\\
	\left.+ {B}_{l-l'}^{(n-n')} \frac{\partial}{\partial \tau} {v_x}_{l'}^{(n'-2)}
	+il'k B_0(x) {v_x}_{l-l'}^{(n-n')} {v_x}_{l'}^{(n')} +B_0(x){v_x}_{l-l'}^{(n-n')}  \frac{\partial}{\partial \xi} {v_x}_{l'}^{(n'-1)} \right] 
	+ \sum_{n',n''=1}^{\infty} \sum_{l',l''=-\infty}^{+\infty}\left[ il''k {B}_{l-l'}^{(n-n')} {v_x}_{l'-l''}^{(n'-n'')} {v_x}_{l''}^{(n'')}	\right.\\
	\left.+ {B}_{l-l'}^{(n-n')}  {v_x}_{l'-l''}^{(n'-n'')} \frac{\partial}{\partial \xi} {v_x}_{l''}^{(n''-1)}\right] 
	+c_s^2\left(\frac{\partial B_0(x)}{\partial x}ilk {B}_{l}^{(n)} +\frac{\partial}{\partial \xi} {B}_{l}^{(n-1)}\right) \\
	+\left[ilk B_0(x) {B}_{l}^{(n)} +B_0(x) \frac{\partial}{\partial \xi} {B}_{l}^{(n-1)} + \sum_{n'=1}^{\infty} \sum_{l'=-\infty}^{+\infty}\left[ il'k {B}_{l-l'}^{(n-n')} {B}_{l'}^{(n')} +{B}_{l-l'}^{(n-n')} \frac{\partial}{\partial \xi} {B}_{l'}^{(n'-1)} +B_0(x) \frac{\partial B_0(x)}{\partial x} +\frac{\partial B_0(x)}{\partial x} B_l^n \right]\right]\\
	+2\Omega_0 \cos \lambda \left[B_0(x) {v_z}_{l}^{(n)} + \sum_{n'=1}^{\infty} \sum_{l'=-\infty}^{+\infty}{B}_{l-l'}^{(n-n')} {v_z}_{l'}^{(n')}\right] - 2\Omega_0 \sin \lambda \left[B_0(x) {v_y}_{l}^{(n)} + \sum_{n'=1}^{\infty} \sum_{l'=-\infty}^{+\infty}{B}_{l-l'}^{(n-n')} {v_y}_{l'}^{(n')}\right]\\
	-\left[B_0(x) \frac{\partial \psi_0(x)}{\partial x} +\frac{\partial \psi_0(x)}{\partial x} B_l^n+ilk B_0(x) {\psi}_{l}^{(n)} + \sum_{n'=1}^{\infty} \sum_{l'=-\infty}^{+\infty}{B}_{l-l'}^{(n-n')} \frac{\partial}{\partial \xi} {\psi}_{l'}^{(n'-1)}+ \sum_{n'=1}^{\infty} \sum_{l'=-\infty}^{+\infty} il'k {B}_{l-l'}^{(n-n')} {\psi}_{l'}^{(n')} \right.\\
	\left. +B_0(x) \frac{\partial}{\partial \xi} {\psi}_{l}^{(n-1)}\right] =0 
\end{split}
\end{equation}
    
\begin{equation}
    \begin{split}
      -i\omega l v_{yl}^{(n)}-\lambda \frac{\partial}{\partial \xi} v_{yl}^{(n-1)}+\frac{\partial}{\partial \tau} v_{yl}^{(n-2)} + \sum_{n'=1}^{\infty} \sum_{l'=-\infty}^{+\infty}\left[ il'k{v_x}_{(l-l'}^{(n-n')} {v_y}_{l'}^{(n')} +{v_x}_{l-l'}^{(n-n')} \frac{\partial}{\partial \xi} {v_y}_{l'}^{(n'-1)} \right] +2\Omega_0 \sin \lambda {v_x}_{l}^{(n)}=0\\  
    \end{split}
\end{equation}
\begin{equation}
    \begin{split}
     -i\omega l v_{zl}^{(n)}-\lambda \frac{\partial}{\partial \xi} v_{zl}^{(n-1)}+\frac{\partial}{\partial \tau} v_{zl}^{(n-2)} + \sum_{n'=1}^{\infty} \sum_{l'=-\infty}^{+\infty}\left[il'k{v_x}_{l-l'}^{(n-n')} {v_z}_{l'}^{(n')} +{v_x}_{l-l'}^{(n-n')} \frac{\partial}{\partial \xi} {v_z}_{l'}^{(n'-1)}\right] -2\Omega_0 \cos \lambda {v_x}_{l}^{(n)}=0\\   
    \end{split}
\end{equation}
    
    \begin{equation}
        \begin{split}
            \frac{\partial^2 \psi_0(x) }{\partial x^2}-l^2k^2\psi^{(n)}_l+2ilk\frac{\partial }{\partial\xi}\psi^{(n-1)}_l +\frac{\partial^2}{\partial \xi^2} \psi^{(n-2)}_l+\omega_J^2 B_0(x)+\omega_J^2 B^{(n)}_l=0\\         
        \end{split}
    \end{equation}
\end{widetext}

\bibliographystyle{elsarticle-num}
\bibliography{ref1}

\begin{thebibliography}{10}
\expandafter\ifx\csname url\endcsname\relax
  \def\url#1{\texttt{#1}}\fi
\expandafter\ifx\csname urlprefix\endcsname\relax\def\urlprefix{URL }\fi
\expandafter\ifx\csname href\endcsname\relax
  \def\href#1#2{#2} \def\path#1{#1}\fi

\bibitem{alfven1942existence}
H.~Alfv{\'e}n, Existence of electromagnetic-hydrodynamic waves, Nature
  150~(3805) (1942) 405--406.

\bibitem{hannan2013fast}
A.~Hannan, T.~Hellsten, T.~Johnson, Fast wave current drive scenarios for demo,
  Nuclear Fusion 53~(4) (2013) 043005.

\bibitem{stasiewicz2000small}
K.~Stasiewicz, P.~Bellan, C.~Chaston, C.~Kletzing, R.~Lysak, J.~Maggs,
  O.~Pokhotelov, C.~Seyler, P.~Shukla, L.~Stenflo, et~al., Small scale
  alfv{\'e}nic structure in the aurora, Space Science Reviews 92 (2000)
  423--533.

\bibitem{stasiewicz2003slow}
K.~Stasiewicz, P.~Shukla, G.~Gustafsson, S.~Buchert, B.~Lavraud, B.~Thid{\'e},
  Z.~Klos, Slow magnetosonic solitons detected by the cluster spacecraft,
  Physical review letters 90~(8) (2003) 085002.

\bibitem{stasiewicz2004theory}
K.~Stasiewicz, Theory and observations of slow-mode solitons in space plasmas,
  Physical review letters 93~(12) (2004) 125004.

\bibitem{shukla2004nonlinear}
P.~K. Shukla, L.~Stenflo, R.~Bingham, B.~Eliasson, Nonlinear effects associated
  with dispersive alfv{\'e}n waves in plasmas, Plasma physics and controlled
  fusion 46~(12B) (2004) B349.

\bibitem{shukla2011alfvenic}
P.~Shukla, B.~Eliasson, L.~Stenflo, Alfv{\'e}nic shock waves in a collisional
  magnetoplasma, Physics Letters A 375~(24) (2011) 2371--2373.

\bibitem{klein2012using}
K.~Klein, G.~Howes, J.~TenBarge, S.~Bale, C.~Chen, C.~Salem, Using synthetic
  spacecraft data to interpret compressible fluctuations in solar wind
  turbulence, The Astrophysical Journal 755~(2) (2012) 159.

\bibitem{schmidt2011slow}
J.~Schmidt, L.~Ofman, Slow magnetoacoustic wave oscillation of an expanding
  coronal loop, The Astrophysical Journal 739~(2) (2011) 75.

\bibitem{rau1998strongly}
B.~Rau, T.~Tajima, Strongly nonlinear magnetosonic waves and ion acceleration,
  Physics of Plasmas 5~(10) (1998) 3575--3580.

\bibitem{marklund2007magnetosonic}
M.~Marklund, B.~Eliasson, P.~K. Shukla, Magnetosonic solitons in a fermionic
  quantum plasma, Physical Review E 76~(6) (2007) 067401.

\bibitem{haas2018magnetosonic}
F.~Haas, S.~Mahmood, Magnetosonic waves in a quantum plasma with arbitrary
  electron degeneracy, Physical Review E 97~(6) (2018) 063206.

\bibitem{masood2014nonlinear}
W.~Masood, R.~Jahangir, B.~Eliasson, M.~Siddiq, A nonlinear model for
  magnetoacoustic waves in dense dissipative plasmas with degenerate electrons,
  Physics of Plasmas 21~(10) (2014).

\bibitem{hussain2011korteweg}
S.~Hussain, S.~Mahmood, Korteweg-de vries burgers equation for magnetosonic
  wave in plasma, Physics of Plasmas 18~(5) (2011).

\bibitem{salahuddin2002ion}
M.~Salahuddin, H.~Saleem, M.~Saddiq, Ion-acoustic envelope solitons in
  electron-positron-ion plasmas, Physical Review E 66~(3) (2002) 036407.

\bibitem{sultana2012electron}
S.~Sultana, I.~Kourakis, Electron-scale electrostatic solitary waves and
  shocks: the role of superthermal electrons, The European Physical Journal D
  66 (2012) 1--12.

\bibitem{mckerr2014freak}
M.~McKerr, I.~Kourakis, F.~Haas, Freak waves and electrostatic wavepacket
  modulation in a quantum electron--positron--ion plasma, Plasma Physics and
  Controlled Fusion 56~(3) (2014) 035007.

\bibitem{chowdhury2017rogue}
N.~Chowdhury, A.~Mannan, A.~Mamun, Rogue waves in space dusty plasmas, Physics
  of Plasmas 24~(11) (2017).

\bibitem{ruderman2010freak}
M.~S. Ruderman, Freak waves in laboratory and space plasmas: Freak waves in
  plasmas, The European Physical Journal Special Topics 185~(1) (2010) 57--66.

\bibitem{habbal1979}
S.~R. Habbal, E.~Leer, T.~E. Holzer,
  \href{https://doi.org/10.1007/BF00151440}{Heating of coronal loops by fast
  mode mhd waves}, Solar Physics 64~(2) (1979) 287--301.
\newblock \href {https://doi.org/10.1007/BF00151440}
  {\path{doi:10.1007/BF00151440}}.
\newline\urlprefix\url{https://doi.org/10.1007/BF00151440}

\bibitem{watanabe1977self}
S.~Watanabe, Self-modulation of a nonlinear ion wave packet, Journal of Plasma
  Physics 17~(3) (1977) 487--501.

\bibitem{sultana2011electrostatic}
S.~Sultana, I.~Kourakis, Electrostatic solitary waves in the presence of excess
  superthermal electrons: modulational instability and envelope soliton modes,
  Plasma Physics and Controlled Fusion 53~(4) (2011) 045003.

\bibitem{shalini2015modulation}
S.~Shalini, N.~Saini, A.~Misra, Modulation of ion-acoustic waves in a
  nonextensive plasma with two-temperature electrons, Physics of Plasmas 22~(9)
  (2015).

\bibitem{misra2007nonlinear}
A.~P. Misra, C.~Bhowmik, Nonlinear wave modulation in a quantum magnetoplasma,
  Physics of plasmas 14~(1) (2007).

\bibitem{bains2010modulational}
A.~Bains, A.~P. Misra, N.~Saini, T.~Gill, Modulational instability of
  ion-acoustic wave envelopes in magnetized quantum electron-positron-ion
  plasmas, Physics of Plasmas 17~(1) (2010).

\bibitem{sahyouni1988dark}
W.~Sahyouni, I.~Zhelyazkov, P.~Nenovski, Dark envelope solitons of fast
  magnetosonic surface waves in solar flux tubes, Solar physics 115 (1988)
  17--32.

\bibitem{sakai1983modulational}
J.-I. Sakai, Modulational instability of fast magnetosonic waves in a solar
  plasma, Solar Physics 84 (1983) 109--118.

\bibitem{misra2008modulational}
A.~Misra, P.~K. Shukla, Modulational instability of magnetosonic waves in a
  spin 1/ 2 quantum plasma, Physics of Plasmas 15~(5) (2008).

\bibitem{panwar2014modulational}
A.~Panwar, C.-M. Ryu, Modulational instability and associated rogue structures
  of slow magnetosonic wave in hall magnetohydrodynamic plasmas, Physics of
  Plasmas 21~(6) (2014).

\bibitem{wang2013modulational}
Y.~Wang, X.~L{\"u}, B.~Eliasson, Modulational instability of spin modified
  quantum magnetosonic waves in fermi-dirac-pauli plasmas, Physics of Plasmas
  20~(11) (2013).

\bibitem{turi2022magnetohydrodynamic}
J.~Turi, A.~Misra, Magnetohydrodynamic instabilities in a self-gravitating
  rotating cosmic plasma, Physica Scripta 97~(12) (2022) 125603.

\bibitem{petrashchuk2023}
A.~V. Petrashchuk, S.~A. Anfinogentov, V.~V. Fedenev, P.~N. Mager, D.~Y.
  Klimushkin, \href{https://doi.org/10.1093/mnras/stad2635}{{Dispersion and
  spatial structure of the coupled Alfvén and slow magnetosonic oscillations
  in the Solar corona}}, Monthly Notices of the Royal Astronomical Society
  525~(4) (2023) 5669--5676.
\newblock \href
  {http://arxiv.org/abs/https://academic.oup.com/mnras/article-pdf/525/4/5669/51607441/stad2635.pdf}
  {\path{arXiv:https://academic.oup.com/mnras/article-pdf/525/4/5669/51607441/stad2635.pdf}},
  \href {https://doi.org/10.1093/mnras/stad2635}
  {\path{doi:10.1093/mnras/stad2635}}.
\newline\urlprefix\url{https://doi.org/10.1093/mnras/stad2635}

\bibitem{ruderman2020}
M.~S. Ruderman,
  \href{https://dx.doi.org/10.1088/1402-4896/aba3a9}{Kadomtsev-petviashvili
  equation for magnetosonic waves in hall plasmas and soliton stability},
  Physica Scripta 95~(9) (2020) 095601.
\newblock \href {https://doi.org/10.1088/1402-4896/aba3a9}
  {\path{doi:10.1088/1402-4896/aba3a9}}.
\newline\urlprefix\url{https://dx.doi.org/10.1088/1402-4896/aba3a9}

\bibitem{andres2017}
N.~Andrés, P.~Clark~di Leoni, P.~D. Mininni, P.~Dmitruk, F.~Sahraoui, W.~H.
  Matthaeus, \href{https://doi.org/10.1063/1.4997990}{{Interplay between
  Alfvén and magnetosonic waves in compressible magnetohydrodynamics
  turbulence}}, Physics of Plasmas 24~(10) (2017) 102314.
\newblock \href
  {http://arxiv.org/abs/https://pubs.aip.org/aip/pop/article-pdf/doi/10.1063/1.4997990/14894464/102314\_1\_online.pdf}
  {\path{arXiv:https://pubs.aip.org/aip/pop/article-pdf/doi/10.1063/1.4997990/14894464/102314\_1\_online.pdf}},
  \href {https://doi.org/10.1063/1.4997990} {\path{doi:10.1063/1.4997990}}.
\newline\urlprefix\url{https://doi.org/10.1063/1.4997990}

\bibitem{asano1969perturbation}
N.~Asano, T.~Taniuti, N.~Yajima, Perturbation method for a nonlinear wave
  modulation. ii, Journal of Mathematical Physics 10~(11) (1969) 2020--2024.

\bibitem{jeans1902stability}
J.~H. Jeans, I. the stability of a spherical nebula, Philosophical Transactions
  of the Royal Society of London. Series A, Containing Papers of a Mathematical
  or Physical Character 199~(312-320) (1902) 1--53.

\bibitem{ichikawa1974c}
Y.~H. Ichikawa, C. nonlinear wave modulation of electrostatic waves, Progress
  of Theoretical Physics Supplement 55 (1974) 212--232.

\bibitem{kolotkov2021}
D.~Y. Kolotkov, D.~I. Zavershinskii, V.~M. Nakariakov,
  \href{https://dx.doi.org/10.1088/1361-6587/ac36a5}{The solar corona as an
  active medium for magnetoacoustic waves}, Plasma Physics and Controlled
  Fusion 63~(12) (2021) 124008.
\newblock \href {https://doi.org/10.1088/1361-6587/ac36a5}
  {\path{doi:10.1088/1361-6587/ac36a5}}.
\newline\urlprefix\url{https://dx.doi.org/10.1088/1361-6587/ac36a5}

\bibitem{ofman2023}
L.~Ofman, T.~A. Kucera, C.~R. DeVore,
  \href{https://dx.doi.org/10.3847/1538-4357/acb13b}{Nonlinear fast
  magnetosonic waves in solar prominence pillars}, The Astrophysical Journal
  944~(2) (2023) 210.
\newblock \href {https://doi.org/10.3847/1538-4357/acb13b}
  {\path{doi:10.3847/1538-4357/acb13b}}.
\newline\urlprefix\url{https://dx.doi.org/10.3847/1538-4357/acb13b}

\bibitem{rust1979}
D.~M. Rust, Z.~{\v{S}}vestka, \href{https://doi.org/10.1007/BF00174535}{Slowly
  moving disturbances in the x-ray corona}, Solar Physics 63~(2) (1979)
  279--295.
\newblock \href {https://doi.org/10.1007/BF00174535}
  {\path{doi:10.1007/BF00174535}}.
\newline\urlprefix\url{https://doi.org/10.1007/BF00174535}

\bibitem{roy2023}
S.~Roy, A.~P. Misra, A.~Abdikian,
  \href{https://doi.org/10.1063/5.0155867}{{Modulation of electromagnetic waves
  in a relativistic degenerate plasma at finite temperature}}, Physics of
  Fluids 35~(6) (2023) 066123.
\newblock \href
  {http://arxiv.org/abs/https://pubs.aip.org/aip/pof/article-pdf/doi/10.1063/5.0155867/18001934/066123\_1\_5.0155867.pdf}
  {\path{arXiv:https://pubs.aip.org/aip/pof/article-pdf/doi/10.1063/5.0155867/18001934/066123\_1\_5.0155867.pdf}},
  \href {https://doi.org/10.1063/5.0155867} {\path{doi:10.1063/5.0155867}}.
\newline\urlprefix\url{https://doi.org/10.1063/5.0155867}

\bibitem{misra2011}
A.~P. Misra, M.~Marklund, G.~Brodin, P.~K. Shukla,
  \href{https://doi.org/10.1063/1.3574913}{{Stability of two-dimensional
  ion-acoustic wave packets in quantum plasmas}}, Physics of Plasmas 18~(4)
  (2011) 042102.
\newblock \href
  {http://arxiv.org/abs/https://pubs.aip.org/aip/pop/article-pdf/doi/10.1063/1.3574913/16718618/042102\_1\_online.pdf}
  {\path{arXiv:https://pubs.aip.org/aip/pop/article-pdf/doi/10.1063/1.3574913/16718618/042102\_1\_online.pdf}},
  \href {https://doi.org/10.1063/1.3574913} {\path{doi:10.1063/1.3574913}}.
\newline\urlprefix\url{https://doi.org/10.1063/1.3574913}

\end{thebibliography}
\end{document}